\documentclass[journal=jacsat,manuscript=article]{achemso}

\usepackage[version=3]{mhchem} 
\usepackage{comment}
\usepackage{multirow}
\usepackage{xcolor}
\usepackage{amsmath}
\usepackage{braket}
\usepackage{amssymb}
\usepackage{relsize}
\usepackage{academicons} 
\usepackage{hyperref}
\usepackage[utf8]{inputenc} 
\usepackage{graphicx}
\usepackage{siunitx}
\usepackage{caption}
\usepackage{subcaption}
\usepackage{geometry}
\usepackage{tocloft}
\usepackage{float}
\usepackage{placeins}
\usepackage{setspace}
\usepackage{booktabs}
\usepackage{lmodern}
\usepackage[T1]{fontenc}

\hypersetup{
    colorlinks=true,
    linkcolor=black,
    citecolor=black,
    urlcolor=black
}

\setkeys{acs}{doi=false}

\newcommand{\orcid}[1]{\href{https://orcid.org/#1}{\textcolor{green}{\aiOrcid}}}

\author{Leonardo Biancorosso}
\affiliation[UniTS]{Dipartimento di Scienze Chimiche e Farmaceutiche, Universit\`a di Trieste, Trieste, Italy}
\alsoaffiliation[ICL]{Department of Materials, Imperial College London, South Kensington Campus, London SW7 2AZ, UK}

\author{Ali Hassanali}
\affiliation[ICTP]{Condensed Matter and Statistical Physics (CMSP), The Abdus Salam Centre for Theoretical Physics, Trieste 34151, Italy}

\author{Mauro Stener}
\affiliation[UniTS]{Dipartimento di Scienze Chimiche e Farmaceutiche, Universit\`a di Trieste, Trieste, Italy}

\author{Emanuele Coccia}
\affiliation[UniTS]{Dipartimento di Scienze Chimiche e Farmaceutiche, Universit\`a di Trieste, Trieste, Italy}
\email{ecoccia@units.it}

\author{Marta Monti}
\affiliation[ICTP]{Condensed Matter and Statistical Physics (CMSP), The Abdus Salam Centre for Theoretical Physics, Trieste 34151, Italy}
\alsoaffiliation[UvA]{Van ’t Hoff Institute for Molecular Sciences, University of Amsterdam, Science Park 904, 1098 XH Amsterdam, The Netherlands}
\email{m.monti@uva.nl}

\author{Gonzalo Díaz Mirón}
\affiliation[ICTP]{Condensed Matter and Statistical Physics (CMSP), The Abdus Salam Centre for Theoretical Physics, Trieste 34151, Italy}
\email{gdiaz_mi@ictp.it}

\title[An \textsf{achemso} demo]
  {Tracking Chirality during Molecular Motor Photoisomerization via Simulated Time-Resolved Circular Dichroism}

\abbreviations{IR,NMR,UV}

\begin{document}

\newpage
\begin{abstract}
 Ultrafast spectroscopic techniques are widely used to investigate photoinduced processes, yet they remain largely blind to molecular chirality. Here we introduce a framework for simulating time-resolved electronic circular dichroism (TRCD) along an ensemble of nonadiabatic molecular dynamics, and apply it to the photoisomerization of a second-generation molecular motor\cite{koumura1999light}. While transient absorption captures the overall excited-state dynamics, it cannot distinguish the two photoproduct pathways. The TRCD response, by contrast, resolves the stereochemical branching: trajectories returning to the stable P isomer (right-handed helix) retain a distinct chiroptical band in the visible region, whereas those forming the M isomer (leftl-handed helix) become chiroptically dark as they twist through the conical intersection. This asymmetry constitutes a directly measurable signature of the stereochemical branching, offering a real-time probe of the formation of molecular chirality and concrete predictions for future TRCD experiments.
\end{abstract}


Light-driven molecular motors convert light energy into controlled mechanical motion and have emerged as versatile building blocks for responsive materials, nanoscale devices, and biomimetic systems\cite{garcia2019light,coskun2012great,van2017dynamic}. In most design, their activity relies on the presence of a stereogenic center that enforces a preferred sense of rotation around a central double bond, enabling unidirectional motion at the molecular scale\cite{roke2018molecular,koumura2002second}. Alternative architectures have nonetheless been reported\cite{greb2014light,wilcken2018complete}, and more recent work achieved directional rotation about single bonds, as well as wavelength-controlled inversion of the rotation direction\cite{nicoli2026wavelength}.

\begin{figure}[H]
    \centering
    \includegraphics[width=0.6\linewidth]{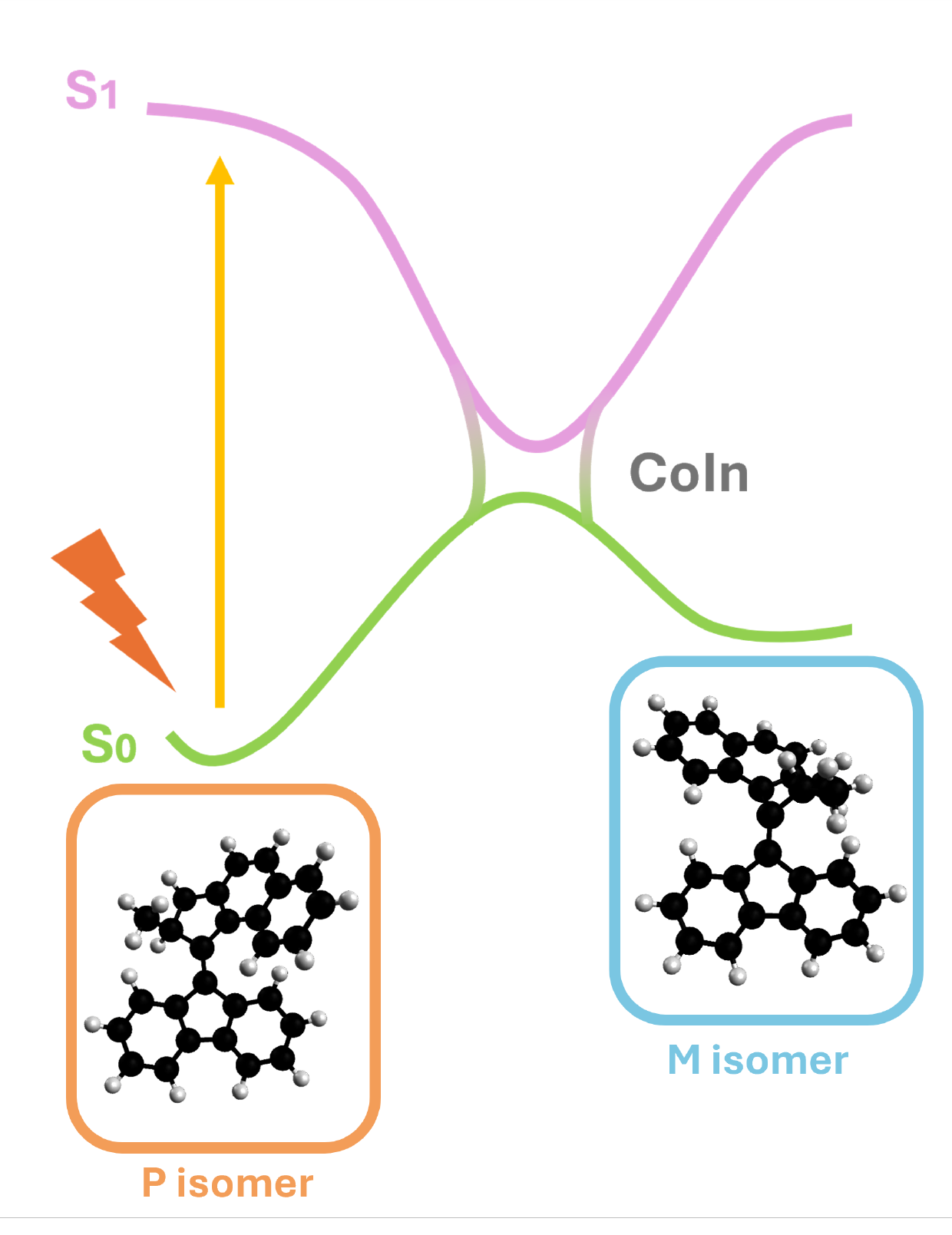}
    \caption{Schematic representation of the photoisomerization process in the second generation motor. Photoexcitation of the P isomer promote the system to the first excited-state $S_1$. The relaxation proceeds through a conical intersection (CoIn), from which the molecule can decay back to $S_0$ yielding either the P or the M isomer.}
    \label{fig1}
\end{figure}

Among the double-bond motors, the second-generation ones introduced by Feringa and co-workers\cite{koumura1999light} are the most extensively investigated. Their rotary cycle consists of a photochemical step followed by a thermally activated process\cite{roy2024excited}. During the photochemical step (Figure \ref{fig1}), absorption of light by the P isomer (right-handed helix) promotes the molecule to the first electronic excited-state ($S_1$), where rotation around the central double bond occurs on ultrafast timescales. Nonradiative relaxation through a conical intersection (CoIn) subsequently returns the system to the electronic ground-state ($S_0$), yielding either the original P isomer or a metastable M conformer (left-handed helix). The M conformer then undergoes a thermal helix inversion that restores the stable P configuration and completes the unidirectional rotation cycle. While considerable effort has been devoted to tuning the thermal step in order to control the rotation rate and improve motor efficiency\cite{tambovtsev2025fine}, the photochemical step remains comparatively less explored despite being the primary event that determines the quantum yield and directionality of motion.

Understanding this photochemical event is challenging because excited-state (ES) relaxation occurs on ultrafast timescales, typically on the order of a picosecond, and involves strongly coupled nuclear and electronic dynamics. Nevertheless, ultrafast spectroscopic studies of molecular motors have provided important insights into these processes. Fluorescence up-conversion measurements revealed an ultrafast transition from an initially bright excited-state to a dark state within the first few hundred femtoseconds after photoexcitation\cite{conyard2012ultrafast,conyard2014chemically}. Complementary transient absorption (TA) experiments further identified the lifetime of this dark state and clarified the timescale of population transfer back to the ground electronic state\cite{hall2017ultrafast,conyard2014chemically}. Together, these studies established a qualitative dynamical picture of the early ES relaxation that drives photoisomerization. Computational investigations have complemented these experiments by providing a more detailed mechanistic description of the excited-state dynamics\cite{gonza2024non,pang2017watching,wen2023excited,filatov2022towards}. In particular, theoretical studies have characterized the topology of the conical intersections responsible for nonadiabatic relaxation and identified vibrational modes that couple to the early-time oscillatory features observed in ultrafast spectroscopic measurements. Both experiment and theory thus describe population transfer and structural evolution in considerable detail. However, neither directly access the evolution of the chiral electronic structure that accompanies the isomerization, a property that is central to the function of the motor. 

In this context, time-resolved circular dichroism (TRCD) spectroscopy\cite{Meyer-Ilse_2013,Hache_2021,monti2024time} represents a particularly powerful probe of ES dynamics in molecular motors. By measuring the transient difference in absorption of left- and right-circularly polarized light, TRCD provides direct sensitivity to changes in the chiral electronic structure of the system. Since the photochemical step of molecular motors involves a substantial evolution of helicity during rotation around the central double bond (P $\rightarrow$ M isomer, Figure \ref{fig1}), TRCD is well suited to follow the associated stereochemical dynamics. Importantly, recent advances in ultrafast pump--probe methodologies have further enabled TRCD measurements on femtosecond timescales, making the technique compatible with the intrinsic time window of ES relaxation\cite{Schmid_2019,Oppermann_2019,Morgenroth_2020,Changenet_2023}.

From the theoretical side, simulations of time-resolved spectra have become increasingly important for connecting ultrafast spectroscopic observables to the underlying nuclear motion\cite{zhang2025understanding, pistillo2026simulating}. In the chiroptical domain, excited-state electronic CD (ESECD) responses have been developed and benchmarked at different levels of electronic-structure theory\cite{scott2021ab}. Theoretical simulations have also been used to interpret ultrafast TRCD measurements, for example in the photoinduced ring-opening of provitamin D, where the time-dependent ground-state (GS) ECD response was connected to the formation of chiral previtamin D rotamers\cite{tapavicza2023}. More directly related to the present work, Padula showed that ESECD can discriminate clockwise and counterclockwise photoisomerization pathways in achiral photo-switches\cite{padula2024discriminating}, providing further support for the sensitivity of probes to stereochemical directionality.

We present a theoretical framework that combines nonadiabatic molecular dynamics (NAMD) with real-time propagation of the electronic wavefunctions to investigate ESECD signals along the photoisomerization pathway of the molecular motor 9-(2-methyl-2,3-dihydro-1H-cyclopenta[a]naphthalen-1-ylidene)-9H-fluorene (see Figure \ref{fig1}).

NAMD simulations were performed using Tight-Binding Time-Dependent Density Functional Theory (TD-DFTB) as implemented in the DFTB+\cite{hourahine2025recent} and SHARC\cite{mai2018nonadiabatic} codes. Additional details can be found in Section S1 and Figures S1--S3 in the Supplementary Information (SI). The employed computational protocol was introduced in a previous study\cite{gonza2024non} and validated against both experimental measurements and higher-level electronic structure methods. To evaluate the time-dependent spectroscopic response, molecular conformations were sampled every 20 fs from a subset of 50 trajectories and used to compute TA and ESECD signals along the ES dynamics. The validation of the selected sub-ensemble is present in Section S1 in the SI.

For each geometry $R_i(t)$, sampled along the $i$-th NAMD trajectory at delay $t$, an electronic dynamics calculation was performed using TDDFT within the Tamm--Dancoff approximation (TDA)\cite{casida,hirata1999time}, employing the CAM-B3LYP functional and the TZP basis set\cite{basis-set}, as implemented in the WaveT code\cite{mon23,biancorosso2024time} interfaced with AMS\cite{ams}. Although TDDFT/TDA is used here, the real-time propagation itself is formulated in a basis of field-free electronic eigenstates and is therefore not tied to this specific electronic-structure level (see Section S2.1 in the SI for theoretical details). In principle, the framework can be combined with higher-level methods, including multiconfigurational approaches and GW/Bethe-Salpeter equation theory,\cite{roos_1980,onida_2002} provided that the required excitation energies and transition moments are available.

The electronic wavefunction, initialized in the NAMD-selected $S_0$ or $S_1$ state, was propagated under a weak impulsive electric-field pulse applied along the three Cartesian directions (FWHM = 94 as; I=$10^2$ W cm$^-2$) to probe the linear optical response. Each calculation yields the time-resolved induced electric and magnetic dipole moments, from which the frequency-resolved polarizability $\bar{\alpha}(\omega)\big|_{R_i(t)}$ and optical-rotation tensor $\bar{\beta}(\omega)\big|_{R_i(t)}$ are obtained. The ensemble-averaged absorption (ABS) and ECD spectra at delay $t$ are then computed as:

\begin{eqnarray}
    \overline{P}^{\text{ABS}}(\omega,t) &=& \frac{4\pi\omega}{c}\,
        \frac{1}{N}\sum_{i=1}^{N}\Im\!\left[\bar{\alpha}(\omega)\big|_{R_i(t)}\right], \\
    \overline{P}^{\text{ECD}}(\omega,t) &=& \frac{4\pi\omega}{c}\,
        \frac{1}{N}\sum_{i=1}^{N}\Im\!\left[\bar{\beta}(\omega)\big|_{R_i(t)}\right],
    \label{spectra1}
\end{eqnarray}
where $c$ is the speed of light, and the sum runs over the $N$ trajectories contributing to the chosen ensemble. Because the NAMD simulations and the optical response calculations rely on different levels of theory, Sections S1--S2 in the SI provides an extensive validation of the computational protocol. The optical response calculations are compared with experimental spectra in Figures S4--S5 in the SI, while Figure S6 assesses the effect of the solvent environment through additional calculations in implicit hexane. These validation tests show that the computed spectra capture the relevant absorption and ECD signatures, and that inclusion of implicit solvent does not alter the conclusions drawn from the vacuum calculations used in the main text.

\begin{figure}[H]
    \centering
    \includegraphics[width=1.0\linewidth]{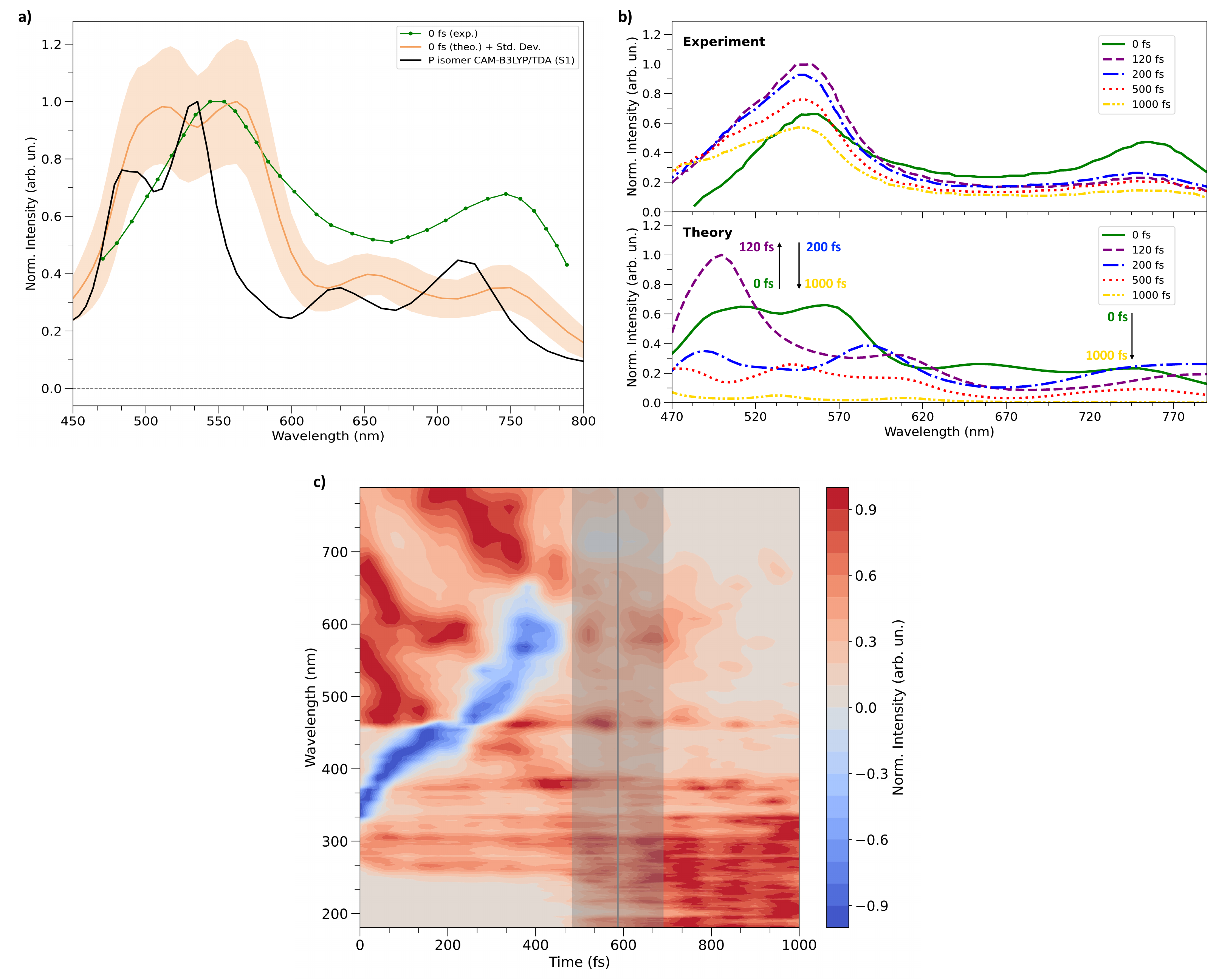}
    \caption{TA response. (a) Experimental (Exp.) excited-state absorption at $t=0$ (green line), compared with the theoretical (theo.) excited-state absorption at $t=0$ averaged over all initial conditions (orange line: mean, orange shaded region: standard deviation) and with the excited-state absorption of the optimized ground-state geometry (black line). (b) Comparison between experimental and calculated transient absorption spectra at representative times along the photodynamics, averaged over the full trajectory ensemble. (c) Two-dimensional transient absorption map calculated along the NAMD trajectories. The gray vertical line indicates the mean hopping time and the shadow region its standard deviation. All experimental data were digitized from Ref.\citenum{hall2017ultrafast}.}
    \label{fig2}
\end{figure}

We begin by analyzing the TA response of the molecular motor along its photodynamics. The results obtained in this work, together with the experimental data from Hall et al.\cite{hall2017ultrafast}, are shown in Figure \ref{fig2}. Transient absorption has proven useful for assigning the characteristic lifetimes of two key steps: the transition from the bright to the dark excited-state, which occurs while the system remains on $S_1$, and the nonradiative $S_1\to S_0$ transition, through which the system returns to the ground-state. In the experimental work (Figure \ref{fig2}a, green line), the band at 760 nm is assigned to the ES absorption from the bright Franck–Condon geometry and shows an ultrafast intensity decay. As the system evolves toward the dark state, a second band grows in at 550 nm on a comparable timescale. Together, these two bands were interpreted as reporting on an equilibrium between the bright and dark states, decaying with an overall lifetime of 1.6 ps as the system returns to the GS. 

Our simulations offer a different interpretation. We find that both bands are already present at the very beginning of the dynamics and can be assigned to excited-state absorption from the bright Franck–Condon geometry itself. Figure \ref{fig2}a shows the average ES absorption computed over the ensemble of geometries at $t = 0$ (orange line). In our calculations, the two bands appear at 530 and 740 nm, in close agreement with the experimental features, and their relative intensity is also reproduced, with the lower-wavelength band being more intense. The same two bands are obtained for the excited-state absorption spectrum computed at the ground-state optimized geometry (Figure \ref{fig2}a, black line), confirming that they are already characteristic of the Franck--Condon structure. We assign these features to the $S_1\to S_{14}$ transition (550 nm band) and to the $S_1\to S_8$ transition (760 nm band). A more extensive transition assignment and corresponding molecular orbitals are reported in Table S2 and Figure S7 in the SI. 

A similar conclusion was reached independently by Wen et al.\cite{wen2023excited} for a related second-generation molecular motor in dimethyl sulfoxide using ADC(2): the absorption around 550 nm, assigned experimentally to a dark state, was found to originate already from the bright Franck--Condon geometry at $t = 0$. In this light, the different experimental interpretation, with the two bands emerging sequentially from bright and dark states, likely reflects the finite time resolution of the measurement (about 100 fs), which is comparable to the timescale of the bright-to-dark-state evolution itself. By contrast, the simulations provide direct access to the initial Franck--Condon region and show that both absorption bands are present from the outset.

Our theoretical spectra reproduce the TA response at early times and capture the main features of its temporal evolution. In the experimental spectra (Figure \ref{fig2}b, top panel), the band around 550 nm increases in intensity at 120 fs (purple line), remains high at 200 fs (blue line), and starts to decrease from 500 fs onward (red line; 1000 fs: yellow line). This growth was assigned to formation of the dark state,\cite{hall2017ultrafast} and our simulations support this assignment: although the band is already present in the bright Franck--Condon state, it gains further intensity as the dark state forms. The calculated spectra (Figure \ref{fig2}b, bottom panel) show the same early enhancement at 120 fs, although the response is blue-shifted and appears as a broader feature, with a strong peak centered near $\sim$500 nm and a weaker contribution near $\sim$600 nm. At 200 and 500 fs, a broad band within the 500--600 nm region is still present, but with substantially reduced intensity, and it becomes negligible by 1000 fs. Thus, the simulations reproduce the overall visible feature and its early rise, while only partially capturing the detailed redistribution of intensity within the 500--600 nm region. The $\sim$750 nm band likewise decreases over time in both experiment and theory, apart from a small experimental blue-shift that is not reproduced by the simulations.

The main discrepancy concerns the timescale of the spectral evolution. Beyond 120 fs, the calculated bands decay faster than in experiment: by 500 fs the computed visible absorption is already strongly reduced, consistent with partial population transfer to S$_0$ within the simulated ensemble ($587.0 \pm 103.4$ fs), and by 1000 fs it has nearly vanished, once all trajectories have relaxed. In contrast, the experimental spectra retain appreciable intensity at both delays. This behavior reflects the shorter ES lifetime obtained in the simulations. The NAMD dynamics, performed in vacuum at the TD-DFTB level, give an $S_1\to S_0$ decay time of about 660 fs,\cite{gonza2024non} compared with the experimental value of 1.6 ps measured in solution.\cite{hall2017ultrafast} In the solvated environment, rotation about the reactive bond that drives the system toward the CoIn is hindered by solvent friction, which lengthens the ES lifetime of related overcrowded-alkene motors.\cite{Harris_2025} This frictional effect is absent in the present vacuum dynamics, leading to faster decay. Together with the approximate ES potential-energy surface, this likely accounts for the underestimated lifetime. The discrepancy therefore concerns the dynamical model rather than the optical-response calculation, which is validated independently in Section S2.

To conclude the analysis of the TA response, Figure \ref{fig2}c presents the two-dimensional spectrum computed over the full trajectory ensemble as a function of time. The two initial bands decrease in intensity as the system approaches the $S_1/S_0$ conical intersection (587.0 $\pm$ 103.4 fs), marked by the gray shaded region in the figure. A third negative feature, assigned to stimulated emission (see Table S2 and Figures S7--S9 in the SI for the molecular orbitals contributions), red-shifts over time as the energy gap between the two potential energy surfaces narrows near the crossing. Once the system has returned to the ground-state through the conical intersection, the bands between 200–300 nm recover, matching the GS absorption features of both the P and M isomers (see Figure S5 in the SI). Analysis of the main transitions and corresponding molecular orbitals at $t = 1000$ fs are shown in Table S2 and Figure S10 in the SI.

As outlined in the introduction, a central goal of this work is to extract isomer-specific information from the simulated spectra. This is not possible with TA in the ground-state, since the relaxed P and M isomers absorb at essentially the same wavelengths\cite{vicario2005controlling}. We therefore examined whether such a distinction might instead emerge in the excited-state, by separating trajectories according to their eventual P or M outcome and computing the corresponding TA response for each subset as a function of time (see Section S1 in the SI for details of this classification). As shown in Figures S11--S14, and discussed in detail in Section S5 in the SI, the two subsets yield nearly identical signals throughout the ES dynamics. This is expected at early times, since all trajectories originate from the P isomer. However, even after reaching the conical intersection, where the pathways have already diverged toward their respective photoproducts (Figure S3 in the SI), the TA signals in the vibrationally hot ground-state remain indistinguishable between the two subsets. This confirms the limitation of linear absorption spectroscopy in following chiral changes\cite{Ress_2023}, whether probed in the ground- or the excited-state.

\begin{figure}[H]
    \centering
    \includegraphics[width=0.8\linewidth]{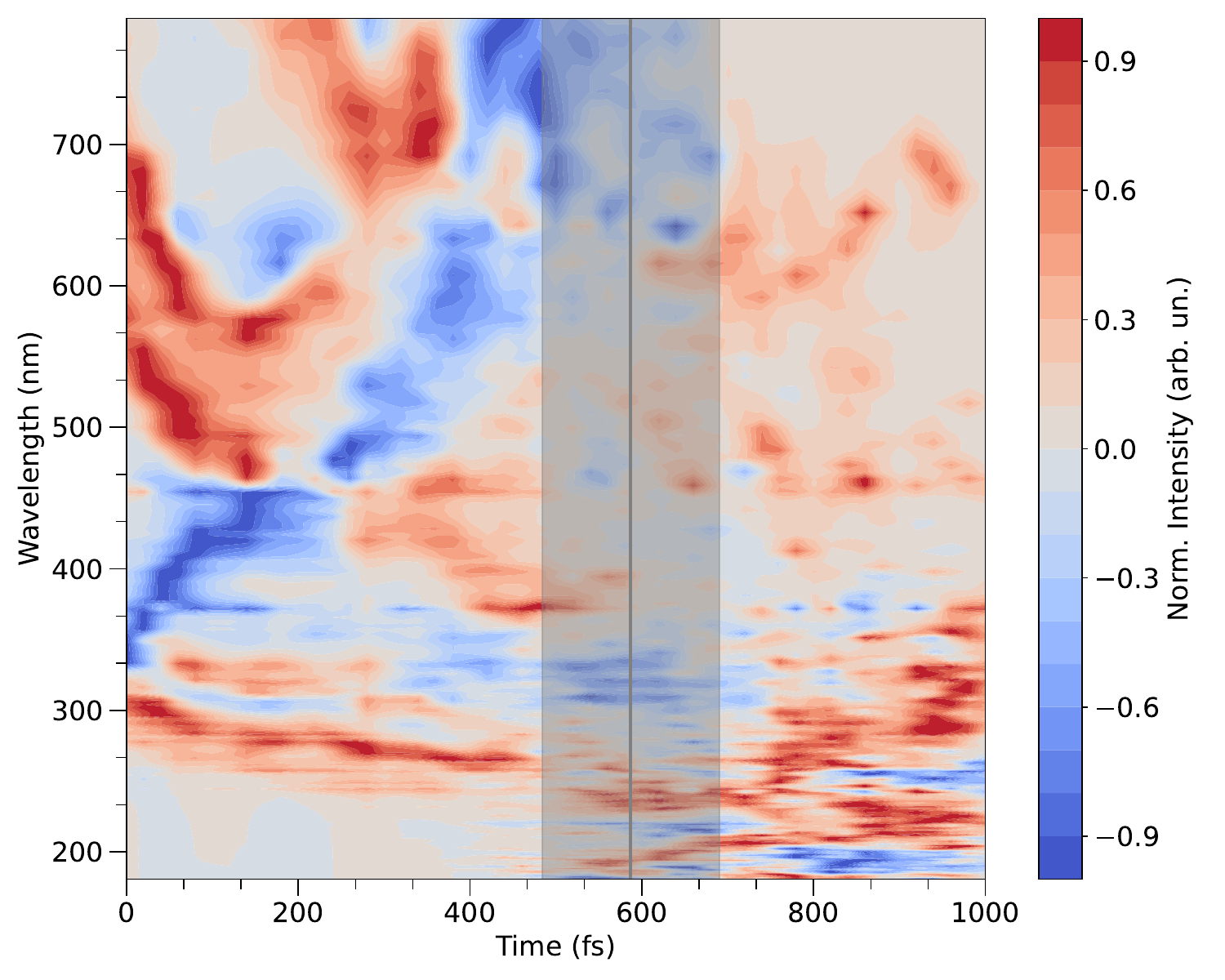}
    \caption{Two-dimensional map of the calculated TRCD response along the NAMD trajectories. The x-axis reports the simulation time at which the molecular geometries were extracted, the y-axis corresponds to the wavelength, and the color scale represents the normalized spectral intensity. The gray vertical line and shaded region indicate the mean $S_1$ $\rightarrow$ $S_0$ decay time and its standard deviation.}
    \label{fig3}
\end{figure}

Having shown that TA cannot resolve the stereochemical identity of the photoproducts, we turn to the calculated TRCD response, which is directly sensitive to the evolving chiral electronic structure. The two-dimensional TRCD spectrum, computed for the full trajectory ensemble, is shown in Figure \ref{fig3}. At $t = 0$ the ES response is characterized by two positive bands in the visible region, centered around 550 and 700 nm, together with a negative band near 320 nm and a positive contribution around 300 nm. These bands originate from the same electronic transitions discussed for the TA spectrum (Table S3 in the SI provides the corresponding assignments for the ECD response).

As the dynamics proceeds, the negative band red-shifts continuously, sweeping across the visible region and overtaking the two early positive bands, which weaken and shift to shorter wavelengths. A further positive contribution appears above 700 nm between 200-350 fs, shortly before the negative band reaches that region. The net effect is a pronounced sign reversal within a single spectral window: around 500–600 nm, the ECD signal is positive at early delays and negative at later ones, directly reflecting the reorganization of the chiral electronic structure as the system approaches the CoIn. A distinct positive band around 300 nm is present throughout the ES dynamics, though its nature changes over time due to the varying weights of the underlying excitations. Beyond the decay window, the visible ECD response becomes weaker and more dispersed, while the strongest features emerge in the UV. In contrast to the TA map, whose ground-state response is uniformly positive, the ECD map shows structured bands of both signs below $\sim$350 nm.

\begin{figure}[H]
    \centering
    \includegraphics[width=1.0\linewidth]{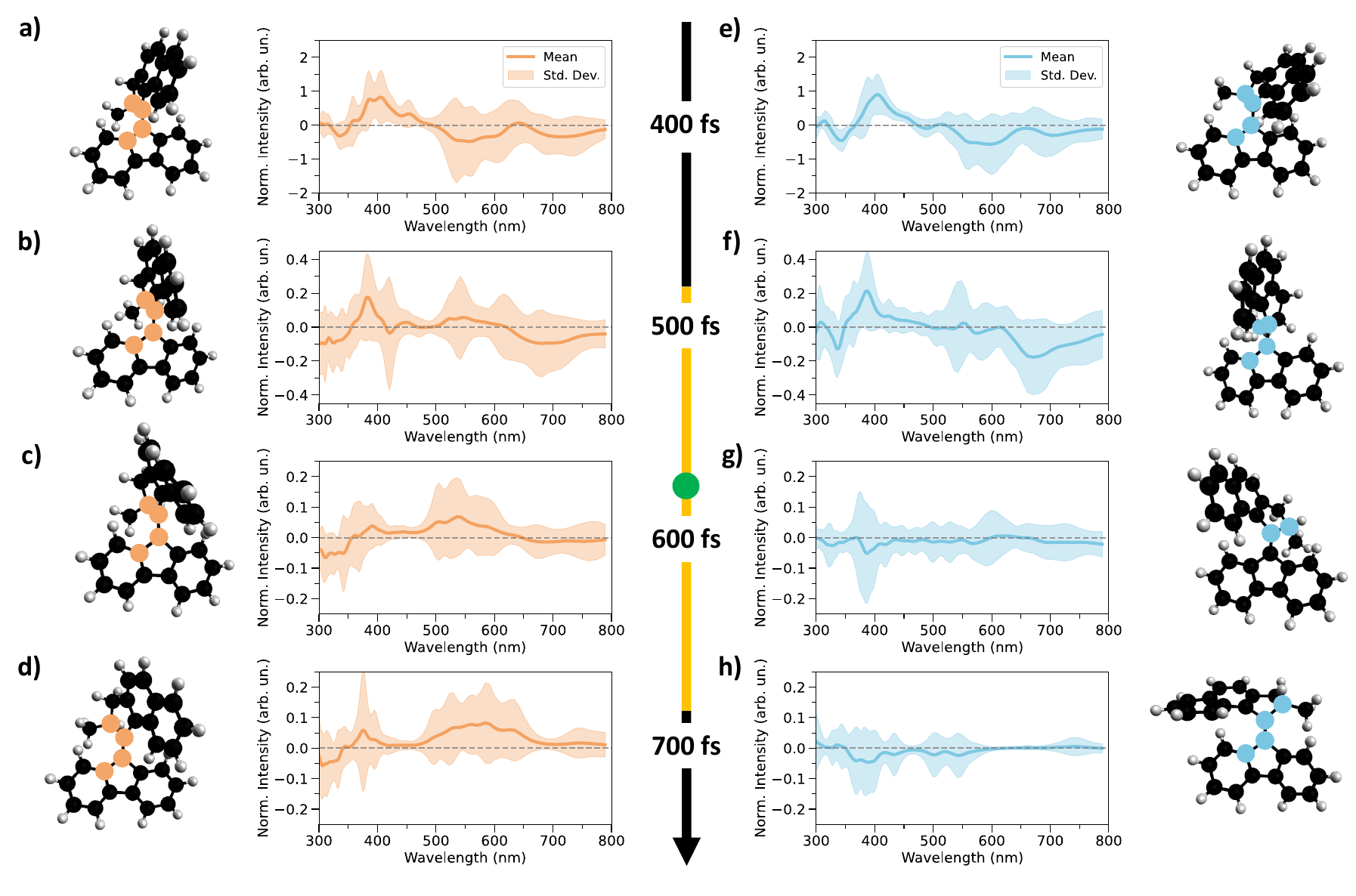}
    \caption{Trajectory-resolved time-dependent electronic circular dichroism (ECD). (a--d) Mean normalized ECD spectra (solid lines) with standard deviation (shaded area) computed along the trajectories associated with the P isomer (orange) with representative molecular structures shown on the left. (e--h) Same for the trajectories associated with the M isomer (blue), with representative structures shown on the right. Spectra are shown in the 300--800 nm range. The central timeline indicates the temporal progression of the dynamics; the green dot marks the mean $S_1$ $\rightarrow$ $S_0$ transition time and the yellow bar the corresponding standard deviation.}
    \label{fig4}
\end{figure}

The ensemble-averaged TRCD spectrum contains no band that can be unambiguously assigned to one isomer over the other: from the experimental observable alone, there is no way to tell whether a given feature originates predominantly from the P or the M isomer. This is precisely where simulation adds diagnostic capability beyond experiment. By classifying trajectories according to their photoproduct identity, we can decompose the ensemble signal into isomer-specific contributions and identify the pathway underlying each spectral feature. Following the same procedure used for the TA response, we compute the trajectory-resolved TRCD spectra separately for the P- and M-bound populations (see Section S1 for the protocol used to assign each trajectory to its corresponding photoproduct). Figure \ref{fig4} shows the resulting spectra for both isomers, focusing on the time and wavelength windows where the isomer-dependent features are most pronounced; the remaining windows are reported in the SI (Figures S15--S17), together with a detailed analysis in Section S7.

During the early stages of the photodynamics, both types of trajectories are chiroptically indistinguishable (Figure \ref{fig4}a--e shows the TRCD at 400 fs, earlier times are given in Figures S15--S16 in the SI). This common response reflects the predominance of a single, P-like configuration at this stage. At 500 fs (panels b and f), as the system begins to transition through the conical intersection (587.0 $\pm$ 103.4 fs), the trajectories start to diverge toward either the P or M isomer (Figure S3 in the SI), and some TRCD features begin to differ accordingly. These differences average out in the ensemble spectrum (Figure \ref{fig3}) and no signal at this stage can be uniquely assigned to either isomer. By 600 fs (panels c and g), differences between the two subsets become clearer, with improved statistics expected to help further disentangle the P- and M-bound signatures. In the visible region (500–-600 nm), where the overall TRCD intensity is approximately one order of magnitude weaker than at 400 fs, P-bound trajectories retain a broad positive band, whereas the M-bound response in this region is largely quenched, fluctuating near zero. This asymmetry mirrors the dihedral evolution (Figure S3 in the SI). At 700 fs (panels d and h), the emerging contrast observed at 600 fs persists and sharpens: the P-bound band around 500–700 nm remains clearly positive and well defined, while the M-bound visible response stays close to zero across the entire range.

Remarkably, this asymmetry persists even at 800 fs (Figure S16, panels c--g in the SI), before the ensemble eventually recovers the signals characteristic of the equilibrated ground-state. This allows us to make the central claim of this work: the positive TRCD band appearing in the 500–700 nm window of the full ensemble-averaged spectrum (Figure \ref{fig3}), emerging at approximately at 600 fs and persisting for roughly 200 fs, can be unambiguously assigned to the P isomer. Crucially, this assignment survives the ensemble average and is therefore, in principle, directly accessible to experiment. The band originates from $S_0\to S_n$ transitions (see Table S3 in the SI) of vibrationally hot ground-state geometries, populated immediately after the trajectories cross the CoIn. These geometries retain a distinct chiroptical signature for roughly 200 fs before cooling. By 1000 fs, the isomer selective feature has disappeared and the TRCD signal reflects the equilibrated ground-state\cite{vicario2005controlling} (Figures S15--S16, panels d--h in the SI). 

In conclusion, we have investigated the photoisomerization of a light-driven molecular motor, with particular focus on how its optical responses reflect the stereochemical evolution between the P and M isomers along the excited-state pathway. The calculated TA spectra reproduce the main experimental features, including the characteristic visible bands and their early-time evolution, supporting the reliability of the simulated excited-state dynamics. Yet absorption, whether experimental or simulated, gives little access to the stereochemical character of the process: the P and M photoproducts remain spectroscopically indistinguishable throughout the pathway. We have shown that TRCD overcomes this limitation. The simulated chiroptical response evolves already within the excited-state--before the two pathways separate in the nuclear coordinates--and is markedly reshaped as the system passes through the conical intersection. Most notably, a positive band in the 500–700 nm window, emerging approximately 600 fs after photoexcitation and persisting for roughly 200 fs, can be unambiguously assigned to the P isomer alone. Crucially, this signature survives averaging over the full trajectory ensemble, and is therefore, in principle, directly measurable: TRCD offers a real-time, isomer-specific fingerprint of the photoisomerization outcome that is inaccessible to transient absorption at any point along the pathway. These results illustrate the value of pairing time-resolved chiroptical spectroscopy with molecular dynamics simulations: while an experimental TRCD measurement necessarily probes the population averaged response, simulation can disentangle it, assigning spectral features to specific geometries, electronic transitions, and photoproduct pathways. This complementarity between experiment and theory is essential to fully exploit TRCD as a stereochemical probe, and the framework introduced here is broadly applicable beyond the present system, offering concrete, testable predictions for future TRCD experiments on molecular motors and other chiral photo-switches.

\begin{acknowledgement}
AH, MM, and GDM acknowledge the European Commission for funding on the ERC Grant HyBOP 101043272.
Financial support from ICSC – Centro Nazionale di Ricerca in High Performance Computing, Big Data and Quantum Computing, funded by European Union – NextGenerationEU is gratefully acknowledged. This work has been supported by the project CHANGE funded by the PRIN 2022 - Progetti di Rilevante Interesse Nazionale (grant 20224KAC28). 
\end{acknowledgement}

\begin{suppinfo}

The Supporting Information reports additional details on the nonadiabatic molecular dynamics simulations, including the computational protocol, validation of the 50-trajectory subset against the full 200-trajectory ensemble, comparison between TD-DFTB and TDDFT/TDA excitation energies, and the trajectory-resolved classification of P and M photoproducts. It also contains the theoretical and computational details of the time-resolved absorption and ECD calculations, validation of the optical response against experimental absorption and ECD spectra, and assessment of solvent effects through vacuum and implicit-hexane calculations. Additional data include transition assignments for transient absorption and TRCD, molecular orbital analyses of the main optical transitions, trajectory-resolved transient absorption spectra, and trajectory-resolved TRCD spectra in both the UV and visible spectral regions.
\end{suppinfo}
\newpage

\newcommand{\D}{\ensuremath{\mathrm{d}}}
\newcommand{\I}{\ensuremath{\mathrm{i}}}

\begin{center}
    \setstretch{2.0}
    {\Large\bfseries\sffamily Supplementary Information for:\\ "Tracking Chirality during Molecular Motor Photoisomerization via Simulated Time-Resolved Circular Dichroism"}
\end{center}

\vspace{0.5em}

\begin{center}
\setstretch{1.5}
\normalsize\sffamily
Leonardo Biancorosso$^{\ddagger,\P}$, Ali Hassanali$^{\mathsection}$, Mauro Stener$^{\ddagger}$, Emanuele Coccia$^{*,\ddagger}$, Marta Monti$^{*,\mathsection,\|}$, and Gonzalo Díaz Mirón$^{*,\mathsection}$
\end{center}


\begin{center}
    \setstretch{1.5}
    \textit{$^{\ddagger}$Dipartimento di Scienze Chimiche e Farmaceutiche, Università di Trieste, Via L. Giorgieri 1, 34127 Trieste, Italy}\\
    \textit{$^{\P}$Department of Materials, Imperial College London, South Kensington Campus, London SW7 2AZ, UK}\\
    \textit{$^{\mathsection}$Condensed Matter and Statistical Physics (CMSP), The Abdus Salam Centre for Theoretical Physics, Trieste 34151, Italy}\\
    \textit{$^{\|}$Van ’t Hoff Institute for Molecular Sciences, University of Amsterdam, Science Park 904, 1098 XH Amsterdam, The Netherlands}
\end{center}

\vspace{0.3em}
\begin{center}
    \sffamily{E-mail:} ecoccia@units.it; m.monti@uva.nl; gdiaz\_mi@ictp.it
\end{center}

\vspace{2em}

\makeatletter
\let\@startsection\acs@startsection
\makeatother
\setcounter{secnumdepth}{3}
\setcounter{table}{0}
\renewcommand{\thetable}{S\arabic{table}}%
\setcounter{page}{1}
\renewcommand{\thepage}{S\arabic{page}}
\setcounter{figure}{0}
\renewcommand{\thefigure}{S\arabic{figure}}%
\setcounter{equation}{0}
\renewcommand{\theequation}{S\arabic{equation}}%
\setcounter{section}{0}
\renewcommand{\thesection}{S\arabic{section}}%
\setcounter{subsection}{0}%
\renewcommand{\thesubsection}{S\arabic{section}.\arabic{subsection}}%

\newpage
\section{Nonadiabatic molecular dynamics simulations}
\subsection{Computational protocol}

As noted in the main text, the NAMD simulations used in this work were presented in our previous study\cite{gonza2024non}. For completeness, however, we summarize here the main features of the protocol. Additional details and validation against higher-level methods and experiment are provided in Ref. \citenum{gonza2024non}.

The workflow begins with geometry optimization in vacuum at the Tight-Binding Density Functional Theory (DFTB) level, followed by frequency calculations to obtain the vibrational normal modes of the optimized structure. From these, 200 initial conditions (atomic positions and velocities) are sampled from the Wigner distribution at 300 K. Each initial condition is then propagated using Tight-Binding Time-Dependent Density Functional Theory (TD-DFTB) within the Tamm–Dancoff approximation, in the NVE ensemble with a 0.5 fs time step, for a total simulation time of 1 ps. Surface hops are identified from the analytical nonadiabatic coupling vectors, and a decoherence correction of 0.1 Hartree was applied for the propagation of the electronic coefficients.

\newpage
\subsection{Ensemble validation}

In this work, we analyze only a subset of the full ensemble presented in our previous study\cite{gonza2024non}. The figures below compare key results obtained for the complete set of 200 trajectories with those from the reduced subset of 50 trajectories analyzed here, and the accompanying table reports the corresponding numerical values.

\begin{figure} [H]
    \centering
    \includegraphics[width=0.7\linewidth]{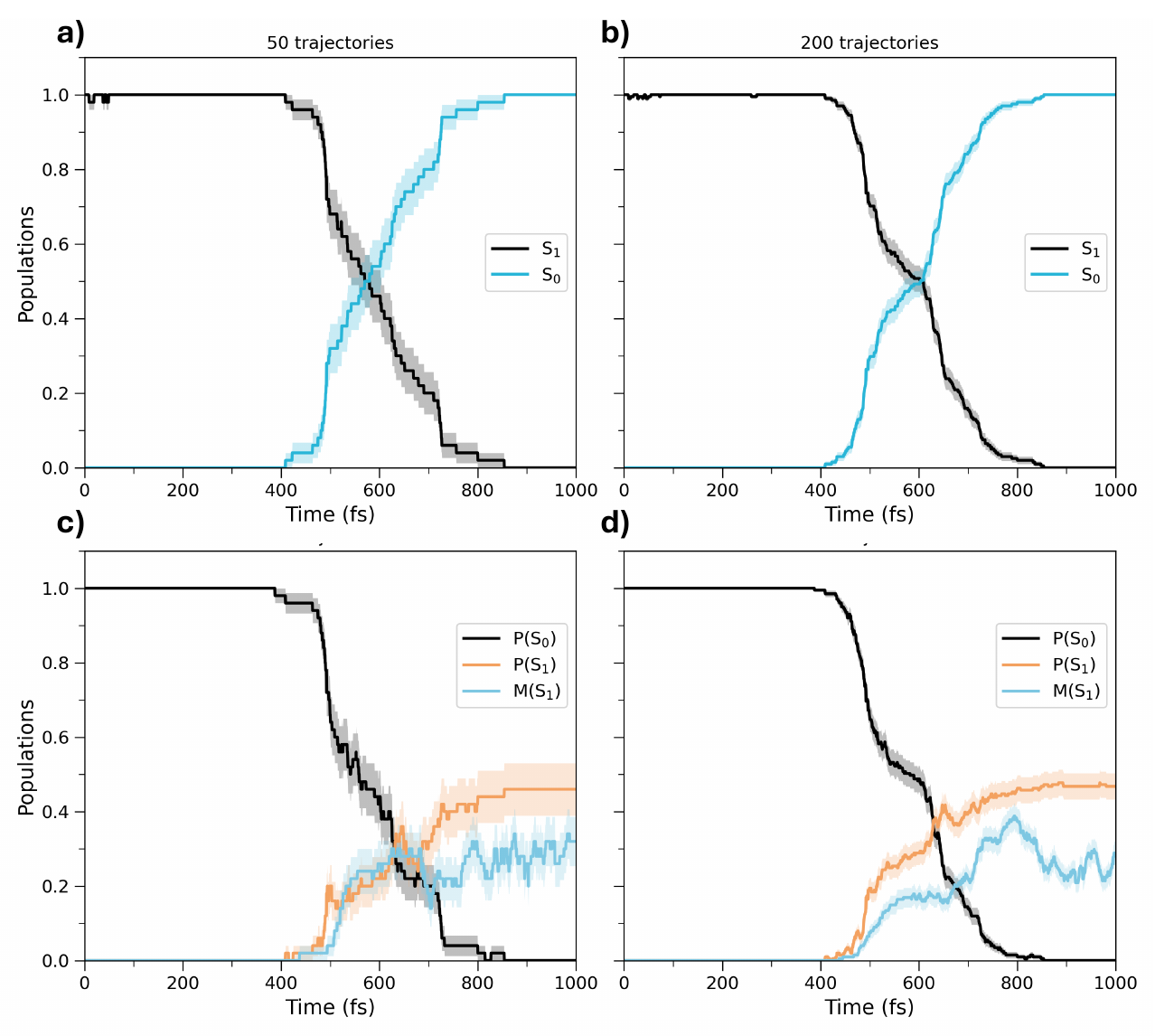}
    \caption{Comparison of the trend in population for the 50 trajectories ensamble and 200 trajectories ensamble. Panel a) and b) show the population trend of the ground state ($S_0$) and the first excited state ($S_1$) respectively in the 50 trajectories ensamble and 200 trajectories ensamble. Panel c) and d) show the populations trend associated with isomer P in the ground state and in the first excited state (P($S_0$) and P($S_1$)) and the isomer M in the first excited state (M($S_1$)).  }
    \label{SI_2}
\end{figure}

\begin{table}[H]
\centering
\caption{Comparison of the 50 and 200 trajectories ensemble.}
\label{tab:my_table}
\begin{tabular}{|c|c|c|}
\hline
Quantity & 50-ensemble (this work) & 200-ensemble (Ref. \citenum{gonza2024non}) \\
\hline
Hopping Time [fs] & 587 $\pm$ 103 & 589 $\pm$ 101 \\
P/M population & 0.46/0.32 & 0.45/0.30 \\
\hline
\end{tabular}
\end{table}

\subsection{Tight-Binding DFT (TD-DFTB) and Time Dependent\break TDDFT/TDA comparison}

\begin{figure}[H]
    \centering
    \includegraphics[width=1.0\linewidth]{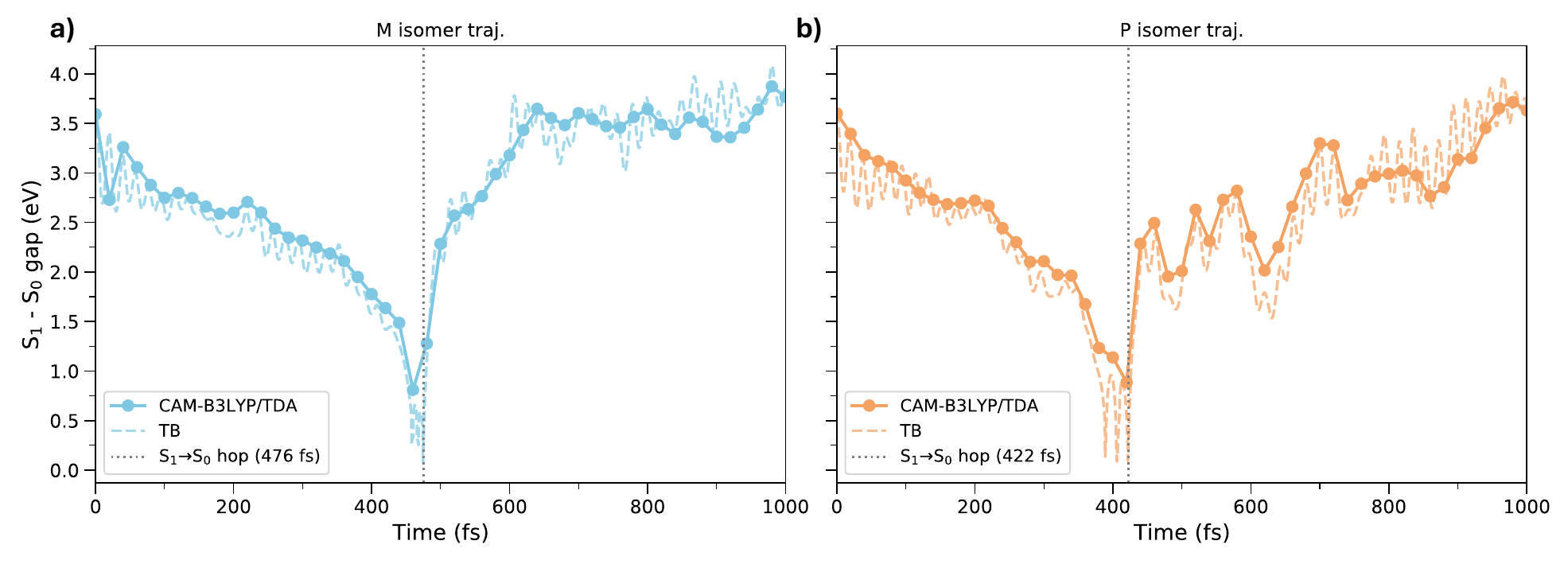}
    \caption{Comparison of $S_0\to S_1$ energy gap using TD-DFTB and TDDFT/TDA (CAM-B3LYP/TZP) for two representative trajectories: (a) evolving toward M isomer, (b) evolving toward the P isomer. Dashed lines are the enegy gap computed using TD-DFTB, extracted from Ref. \citenum{gonza2024non}. Energy gap at TDDFT level were computed at geometries sampled every 20 fs and they are represented as dotted lines.}
    \label{fig:comparison_TD_TB}
\end{figure}

Because our protocol combines TD-DFTB for the photodynamics with TDDFT/CAM-B3LYP under the TDA for the optical response, we first verify that the two levels of theory yield consistent excitation energies along representative trajectories. Figure \ref{fig:comparison_TD_TB} compares the $S_0\to S_1$ energy gap computed with both methods along two representative trajectories, one evolving toward the P isomer and the other toward the M isomer.

\subsection{Trajectory-Resolved Classification of P and M Photoproducts}

\begin{figure}[H]
    \centering
    \includegraphics[width=0.6\linewidth]{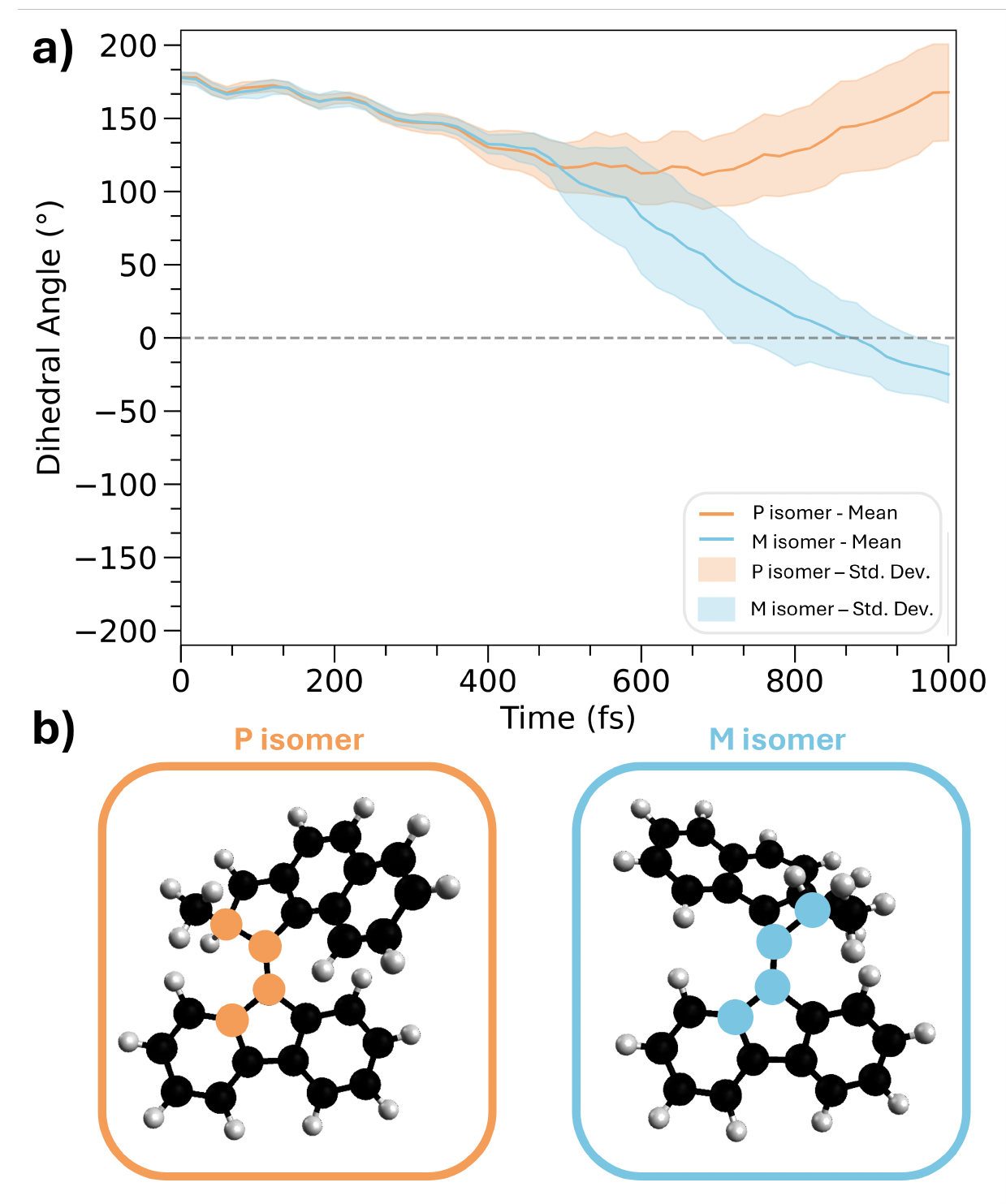}
    \caption{(a) Time evolution of the mean dihedral angle for P (orange) and M (blue) trajectories, with the corresponding standard deviations shown as shaded areas. (b) Representative final structures at 1000 fs for the two sets of trajectories: P isomer on the left and M isomer on the right. The atoms defining the tracked dihedral angle are highlighted in orange and blue, respectively}
    \label{SI_dih}
\end{figure}

In this work, the classification of trajectories as leading to the M or P isomer is based on the dihedral angle of the final structure, defined by the atoms highlighted in Figure \ref{SI_dih}. If the final dihedral angle is smaller than 100°, the photoisomerization process is considered to have occurred, and the final structure is classified as a possible configuration of the M isomer. Conversely, if the final dihedral angle is larger than 100°, the final structure is classified as a possible configuration of the P isomer, and the corresponding trajectory is accordingly classified as leading to the P isomer. Using this classification, in the subset of 50 trajectories, 26 leads to the formation of the M isomer and 24 brings to the formation of the P isomer. 

\newpage

\section{Time-resolved electronic circular dichroism spectra}
\subsection{Theoretical details}
\label{theo}

Absorption (ABS) and electronic circular dichroism (ECD) spectra were computed through real-time electronic dynamics based on the time-dependent Schrödinger equation (TDSE):

\begin{equation}
    i \frac{d}{dt} \ket{\Psi(t)} = \hat{H}(t) \ket{\Psi(t)} ,
    \label{TDSE}
\end{equation}
where the Hamiltonian is partitioned into a field-free and a field-interaction term:

\begin{equation}
    \hat{H}(t) = \hat{H}_{0} + \hat{H}_{F}(t).
    \label{Eq. Ham}
\end{equation}
The interaction term in Eq. \ref{Eq. Ham}, expressed in the length gauge, becomes:

\begin{equation}
    \hat{H}_{F}(t) = - \vec{\hat{\mu}} \cdot \vec{F}(t), 
\end{equation}
with $\vec{\hat{\mu}}$ being the electric dipole operator and $\vec{F}(t)$ the external electric field, applied as a weak impulsive perturbation to probe the linear response.

The time-dependent wavefunction $\ket{\Psi(t)}$ is expressed as a linear combination of the $N_{states}$ eigenstates $\ket{\lambda}$ of the field-free Hamiltonian $\hat{H}_0$--namely the ground-state (GS) and the $N_{states}$-1 time-dependent density functional theory (TDDFT) excited-states:

\begin{equation}
    \ket{\Psi(t)} = \sum_{\lambda=0}^{N_{states}-1} C_{\lambda}(t) \ket{\lambda},
    \label{wavefunc}
\end{equation}
where $C_{\lambda}(t)$ are the time-dependence coefficients. In this space, Eq. \ref{TDSE} can be recast in matrix form:

\begin{equation}
    i \frac{d \mathbf{C}(t)}{dt} = \mathbf{H}(t)\mathbf{C}(t),
\end{equation}
where $\mathbf{C}(t)$ is the vector of expansion coefficients and $\mathbf{H}(t)$ is the matrix representation of the time-dependent Hamiltonian at time $t$ . 
The electronic states were described within a singly-excited configuration ansatz:

\begin{equation}
|\lambda\rangle = d_0|\psi_0\rangle + \sum_{i}^{\text{occ}}\sum_{a}^{\text{vir}} d_{i,\lambda}^{a}|\psi_i^a\rangle ,
\end{equation}
where $|\psi_0\rangle$ and $|\psi_i^a\rangle$ are the GS Slater determinant and the configuration obtained by singly exciting from the occupied molecular orbital $i$ to the virtual molecular orbital $a$, respectively, while $d_0$ and $d_{i,\lambda}^{a}$ are the expansion coefficients. In the present work, this ansatz is used within the Casida formulation of TDDFT. Time propagation was performed using a second-order Euler scheme \cite{pipolo16,ec18}.

For each sampled frame (see the Main Text), the electronic wavefunction (Eq. \ref{wavefunc}) was initialized in the state occupied by that trajectory at that time: a frame propagating on $S_1$ was initialized as $|C_{1}(t=0)|^2=1$, and one on $S_0$ as $|C_{0}(t=0)|^2=1$. This pure-state initial condition, set frame by frame by the active surface of the surface-hopping dynamics, provides the starting point for the real-time propagation, from which the induced electric $(\Delta\vec{\mu}(t))$ and magnetic $(\Delta\vec{m}(t))$ dipoles are obtained:

\begin{equation}
    \Delta\vec{\mu}(t)=\vec{\mu}(t)-\vec{\mu}(0),
    \label{eq:ineldip}
\end{equation}
and

\begin{equation}
    \Delta\vec{m}(t)=\vec{m}(t)-\vec{m}(0).
    \label{eq:inmagdip}
\end{equation}
$\vec{\mu}(t)$ and $\vec{m}(t)$ in Eqs. \ref{eq:ineldip}--\ref{eq:inmagdip} correspond to the time-dependent electric and magnetic moments at time $t$, while $\vec{\mu}(0)$ and $\vec{m}(0)$ to the same quantities but at $t=0$. Both quantities are computed from the TDSE propagation in state space: 
\begin{equation}
\vec{\mu}(t) = \sum_{\lambda'\lambda} C_{\lambda'}^{*}(t)C_{\lambda}(t)
\bra{\lambda'} \vec{\hat{\mu}} \ket{\lambda},
\end{equation}
and

\begin{equation}
\vec{m}(t) = \sum_{\lambda'\lambda} C_{\lambda'}^{*}(t)C_{\lambda}(t)
\bra{\lambda'} \vec{\hat{m}} \ket{\lambda}.
\end{equation}
where $\bra{\lambda'}\vec{\hat{\mu}}\ket{\lambda}$ and $\bra{\lambda'}\vec{\hat{m}}\ket{\lambda}$ are the transition electric and magnetic moments, respectively, and $\lambda'$ and $\lambda$ are the electronic states. \\
Both elements from the GS to an excited $\lambda>0$ state, and elements in which both $\lambda$ and $\lambda'$ refer to an ES are computed. 
Derivation of these transition moments can be found in Refs. \cite{mon23,biancorosso2024time}

The rotationally-averaged ABS and ECD spectra are then calculated as:

\begin{eqnarray}
    P^{\text{ABS}} & = & \frac{4 \pi \omega}{c} \Im[\bar{\alpha}(\omega)] \\
     P^{\text{ECD}} & = & \frac{4 \pi \omega}{c} \Im[\bar{\beta}(\omega)], 
\end{eqnarray}
where $c$ is the light speed and 
\begin{equation}
\bar{\alpha}(\omega) = \frac{1}{3} (\alpha_{xx} + \alpha_{yy} + \alpha_{zz}), 
\label{eq:abs}
\end{equation}
\begin{equation}
\bar{\beta}(\omega) = \frac{1}{3} (\beta_{xx} + \beta_{yy} + \beta_{zz}), 
\label{eq:rot}
\end{equation}
and 

\begin{eqnarray}
    \alpha_{nl}(\omega) & = & \frac{1}{2\pi F_n(\omega)} \int_0^{+\infty} \Delta {\mu}_l(t)\, e^{i(\omega+i\gamma)t}\, dt, \\
       \beta_{nl}(\omega) & = & -\frac{i}{2\pi\omega {F}_{n}(\omega)}\int_{0}^{+\infty}-\Delta m_l(t)e^{i(\omega+i\gamma)t}dt,
    \label{eq:ABS_spc}
\end{eqnarray}
where $\Delta{\mu}_l(t)$ and $\Delta{m}_l(t)$ represent the $l$th component of the induced electric and magnetic dipole moment, respectively \cite{pis26,mon23,biancorosso2024time}, $F_n(\omega)$ is the Fourier transform of the $n$th component of the external electric field, and  $\gamma$ is a damping parameter that is used to describe the ES lifetime. 

\subsection{Computational protocol}
\label{det_ESECD}
For the calculation of the transient absorption (TA) and excited-state electronic circular dichroism (ESECD) spectra, 50 NAMD trajectories were randomly selected from the full ensemble, and molecular structures were extracted from each trajectory every 20 fs. At each extracted geometry, a frequency-domain TDDFT calculation was performed within the Tamm-Dancoff approximation (TDA)\cite{hirata1999time} using the Casida formalism,\cite{casida} as implemented in the Amsterdam Modeling Suite (AMS) software\cite{ams}. The calculations employed the CAM-B3LYP\cite{YANAI200451} functional and a triple-zeta plus polarization (TZP) basis set of Slater-type orbitals\cite{basis-set}, retaining the lowest 100 excited-states. Excitation energies and transition electric and magnetic dipole moments were extracted from these calculations through an in-house interface\cite{grobas_2021,mon23,biancorosso2024time} between AMS and the WaveT package,\cite{pipolo16} and used as input parameters for the subsequent real-time propagation. 

Real-time electronic dynamics were then propagated for 100 fs with a timestep of 1 as, starting from the electronic state occupied by the trajectory at the extracted time (see Section \ref{theo}). The external field was applied along the three Cartesian directions and the spectra were calculated as described in Section \ref{theo}. The field parameters correspond to a pulse with FWHM = 94 as and intensity $I = 10^{2}$ W cm$^{-2}$.

\subsection{Optical response validation with experiments}
The optical response of the two isomers was validated against experimental data prior to the time-resolved analysis. As mentioned in Section \ref{det_ESECD}, absorption and ECD spectra were computed within the TDA, consistent with the level of theory adopted for the ES description in the NAMD calculations\cite{gonza2024non} and further motivated by the well-established ability of TDA to mitigate numerical instabilities near conical intersections\cite{hu2014performance,li2014configuration}.

\begin{figure}[H]
    \centering
    \includegraphics[width=0.8\linewidth]{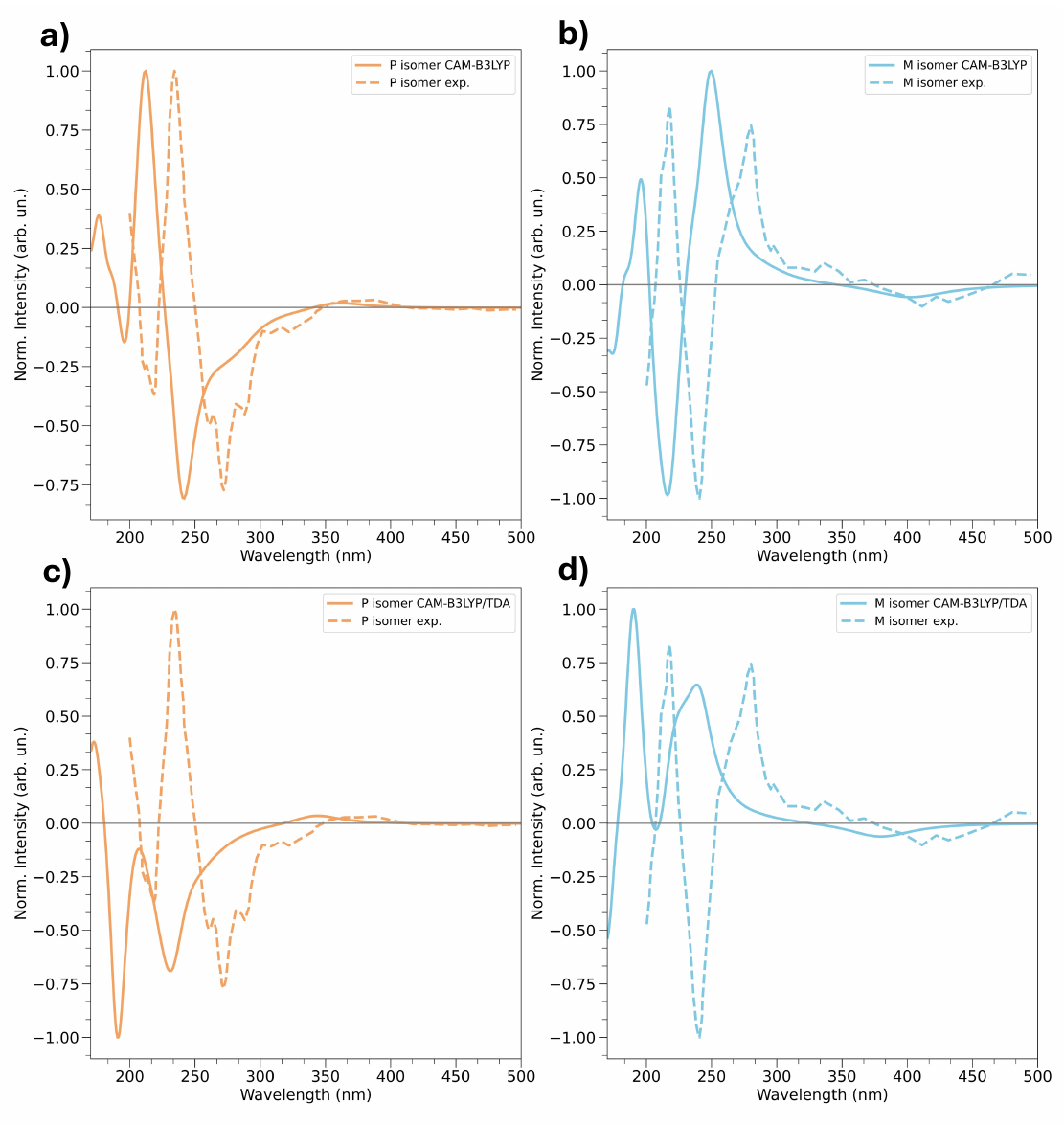}
    \caption{Experimental spectra (dashed lines; P: orange; M: blue) are compared with spectra calculated at the CAM-B3LYP level in panels (a) and (b), and at the CAM-B3LYP/TDA level in panels (c) and (d), for the P and M isomers, respectively. Experimental data were taken from Ref. \citenum{vicario2005controlling}. All spectra are normalized to their maximum absolute intensity.}
    \label{SI_4}
\end{figure}

\begin{figure}[H]
    \centering
    \includegraphics[width=1.0\linewidth]{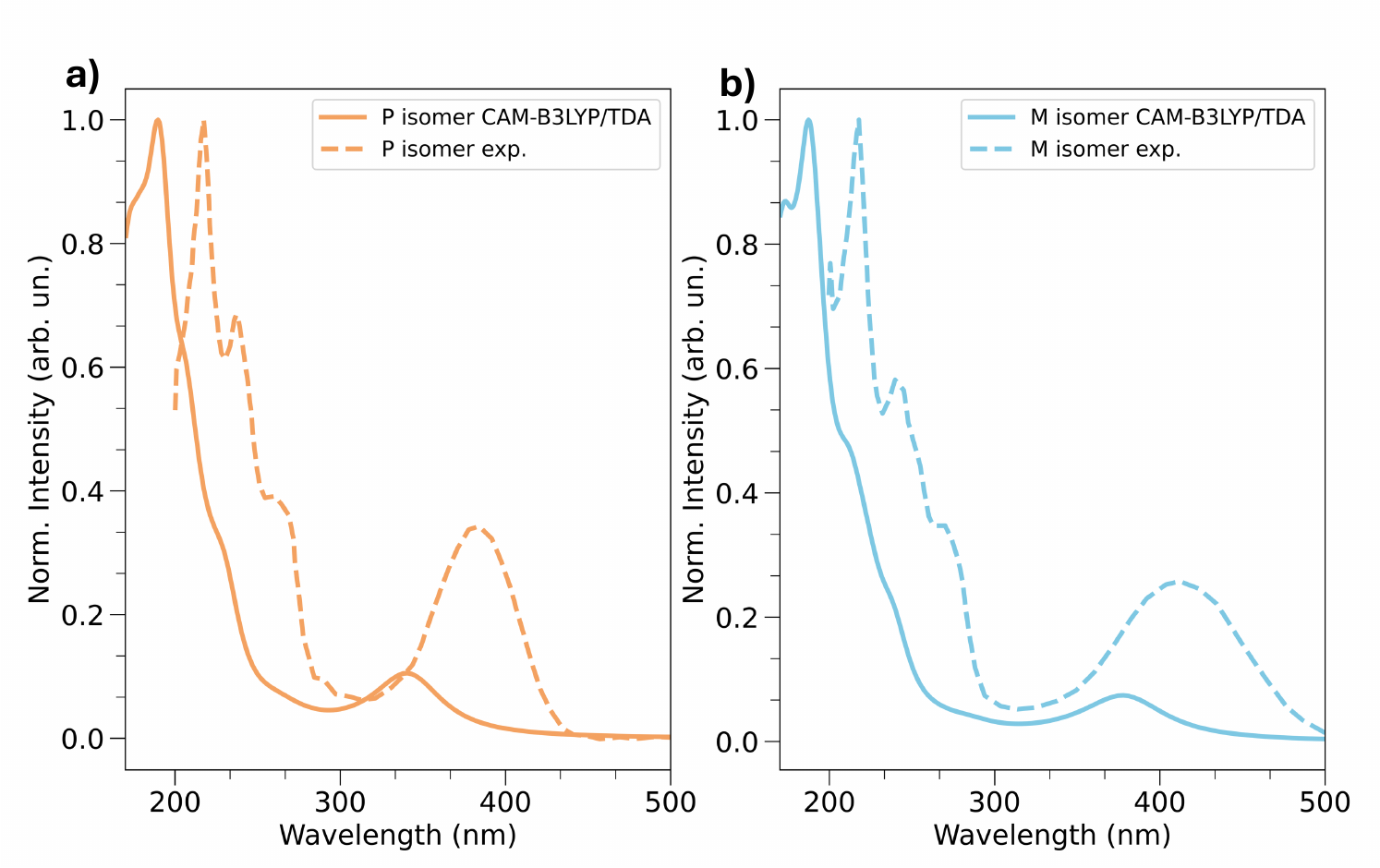}
    \caption{Comparison between the calculated (solid lines) and experimental (dashed lines) absorption spectra of the P and M isomers in the ground state, shown in panels (a) and (b), respectively.}
    \label{SI_5}
\end{figure}

Nevertheless, we assessed the quality of the approximation by comparing the full TDDFT and TDA spectra of the ground-state optimized isomers against experimental results\cite{vicario2005controlling}. The ECD comparison is shown in Figure \ref{SI_4}. Both calculations are blue-shifted with respect to experiment, although the full TDDFT spectra (panels a--b) reproduce the relative intensity of the main features more accurately than TDA. The TDA approximation introduces additional differences in peak positions and relative intensities, but still preserves the signed structure of the response and clearly distinguishes the P and M isomers (panels c--d). Absorption is captured well at the TDA level (Figure \ref{SI_5}): the two main bands are reproduced for both isomers, with only a modest blue-shift with respect to experiment and a slight underestimation of the intensity of the band at longer wavelength. We therefore adopt TDA for the time-resolved simulations, as it provides a qualitatively reliable chiroptical response while offering improved numerical stability for the distorted geometries sampled along the nonadiabatic dynamics.

\begin{figure}[H]
    \centering
    \includegraphics[width=1.0\linewidth]{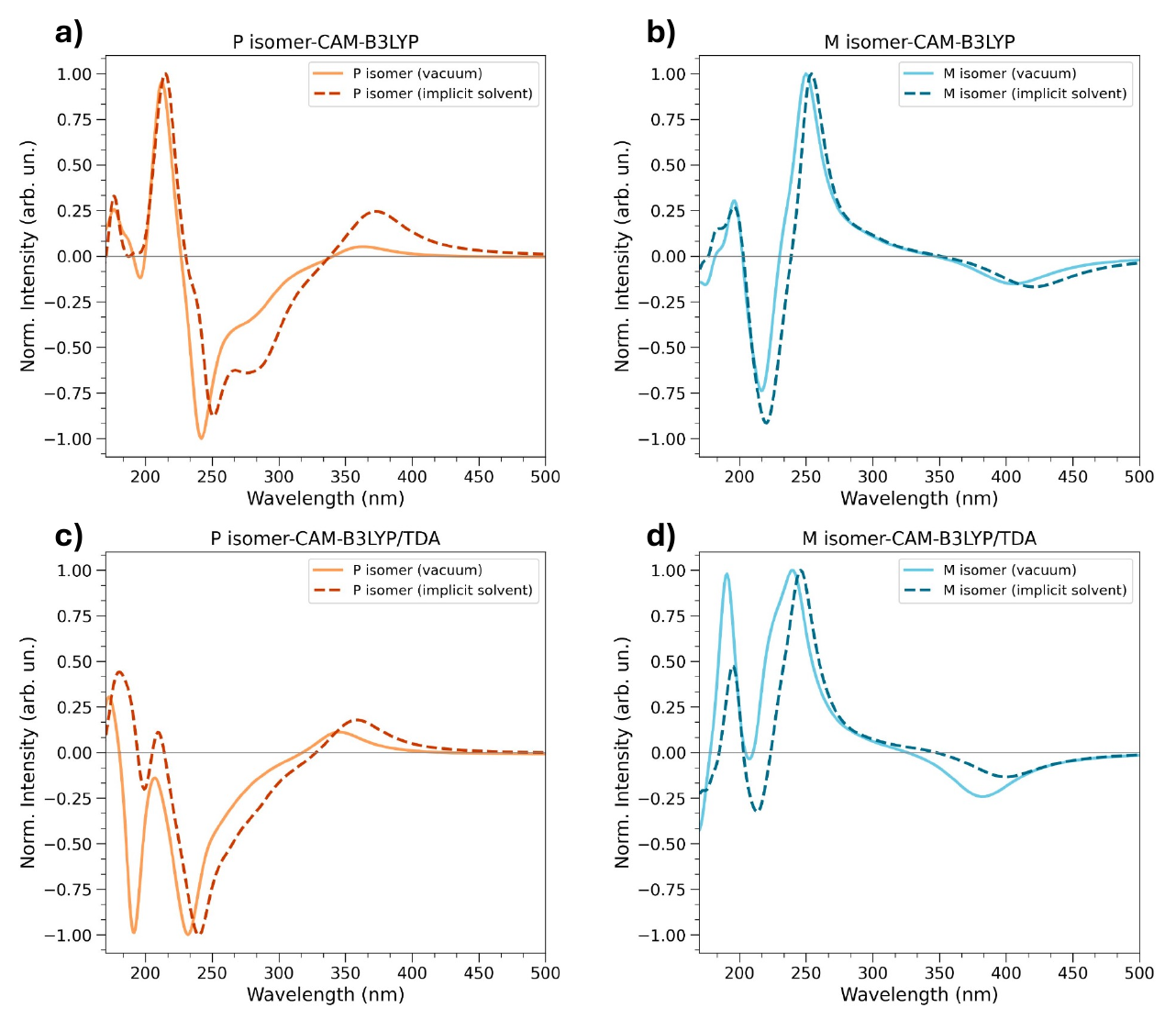}
    \caption{Comparison of ECD spectra computed in vacuum and with implicit solvent (hexane, COSMO) for the P and M isomers. Panels (a) and (b) show results obtained at the CAM-B3LYP level for the P and M isomers, respectively, while panels (c) and (d) report the corresponding spectra computed using the TDA approximation. Solid lines represent gas-phase calculations, and dashed lines include implicit solvent effects.}
    \label{SI_6}
\end{figure}

Both full TDDFT and TDA spectra were computed in vacuum, whereas the experimental spectra were measured in hexane. To assess the possible influence of the solvent environment, additional calculations were performed using hexane as an implicit solvent within the COSMO scheme\cite{Klamt_1993}, consistent with the experimental conditions\cite{vicario2005controlling}. As shown in Figure \ref{SI_6}, inclusion of the nonpolar solvent leads only to minor changes in the spectral shape and relative intensities, without altering the overall chiroptical response. This is consistent with the generally weak influence of nonpolar solvents on ECD spectra\cite{Pescitelli_2016,Monti_2023}. Therefore, all results discussed in the main text are reported for calculations performed in vacuum.

\section{Optical transitions in transient absorption}

\begin{table}[H]
\centering
\caption{Transition strengths computed along the nonadiabatic dynamics at 0, 120, 200, 500, and 1000 fs, within the 450–800 nm spectral window. For each selected time, we report the most relevant electronic transitions contributing to the absorption spectrum. For each transition, the average wavelength $\overline{\lambda}$ and transition strength $\overline{D}$ are reported together with their standard deviations, obtained from the ensemble of trajectories. The transition strength is defined as $D = (\mu_x^2 + \mu_y^2 + \mu_z^2)/3$.}
\begin{tabular}{|c|l|c|c|c|c|}
\hline
\rule{0pt}{2.8ex}\textbf{Time (fs)} & \textbf{Transition} &
\textbf{$\overline{\lambda}$ $\pm \; \sigma_\lambda$ (nm)} &
\textbf{$\overline{D} \pm \sigma_D$} \\
\hline
\multirow{4}{*}{0}
 & $|1\rangle \to |0\rangle$   & 344 $\pm$ 6 &  -2.7 $\pm$ 0.1 \\
 & $|1\rangle \to |17\rangle$  & 516 $\pm$ 9 & 0.4 $\pm$ 0.2   \\
 & $|1\rangle \to |14\rangle$  & 567 $\pm$ 9 & 0.8 $\pm$ 0.2   \\
 & $|1\rangle \to |8\rangle$   & 754 $\pm$ 17 & 0.9 $\pm$ 0.1   \\
\hline
\multirow{4}{*}{120}
 & $|1\rangle \to |0\rangle$   & 427 $\pm$  12 & -3.2 $\pm$ 0.2   \\
 & $|1\rangle \to |12\rangle$  & 501 $\pm$  10 & 0.9 $\pm$ 0.5  \\
 & $|1\rangle \to |8\rangle$   & 608 $\pm$  14 & 0.8 $\pm$ 0.1  \\
 & $|1\rangle \to |7\rangle$   & 768 $\pm$  26 & 0.6 $\pm$ 0.2  \\
\hline
\multirow{4}{*}{200}
 & $|1\rangle \to |0\rangle$   & 445 $\pm$  16 & -3.0 $\pm$ 0.2  \\
 & $|1\rangle \to |14\rangle$  & 474 $\pm$  12 & 0.4 $\pm$ 0.3  \\
 & $|1\rangle \to |9\rangle$   & 589 $\pm$  14 & 1.0 $\pm$ 0.2  \\
 & $|1\rangle \to |7\rangle$   & 744 $\pm$  26 & 0.6 $\pm$ 0.3  \\
\hline
\multirow{4}{*}{500}
 & $|1\rangle \to |0\rangle$   & 835 $\pm$ 285 & -2.0 $\pm$ 0.3  \\
 & $|1\rangle \to |8\rangle$   & 507 $\pm$ 54 & 0.03 $\pm$ 0.04  \\
 & $|1\rangle \to |6\rangle$   & 616 $\pm$  83 & 0.5 $\pm$ 0.4  \\
 & $|1\rangle \to |3\rangle$   & 760 $\pm$ 125 & 0.9 $\pm$ 0.2  \\
\hline
\multirow{4}{*}{1000}
 & $|0\rangle \to |18\rangle$  & 211 $\pm$   5 & 0.4 $\pm$ 0.4  \\
 & $|0\rangle \to |15\rangle$  & 218 $\pm$   6 & 0.4 $\pm$ 0.4  \\
 & $|0\rangle \to |14\rangle$  & 221 $\pm$   6 & 0.4 $\pm$ 0.4  \\
 & $|0\rangle \to |1\rangle$   & 370 $\pm$  56 & 1.9 $\pm$ 0.9  \\
\hline
\end{tabular}
\vspace{0.5cm}
\end{table}

\section{Molecular Orbitals for the Optical responses along time}

\begin{figure}[H]
    \centering
    \includegraphics[width=1.0\linewidth]{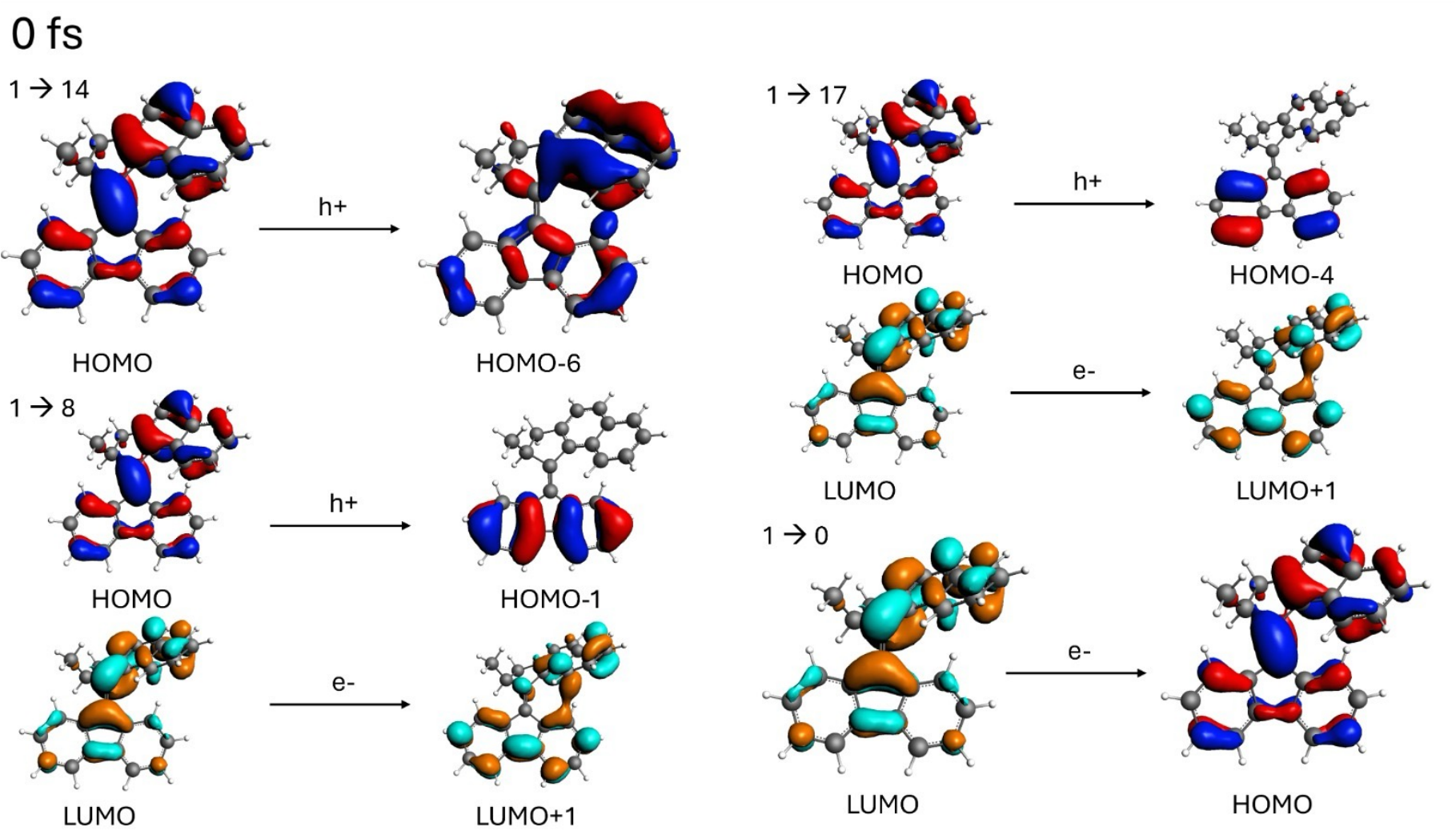}
    \caption{Isosurfaces of the frontier molecular orbitals involved in the four representative hole (h$^+$) and electron (e$^-$) transitions at 0 fs, labeled by the corresponding excitation (1 $\rightarrow$ 14, 1 $\rightarrow$ 8, 1 $\rightarrow$ 17, 1 $\rightarrow$ 0). Occupied orbitals (HOMO, HOMO-1, HOMO-4, HOMO-6) are shown in blue and red, denoting the two opposite phases of the wavefunction; unoccupied orbitals (LUMO, LUMO+1) are shown in orange and cyan, again denoting opposite phases. Arrows indicate the orbital-to-orbital character of each hole or electron transfer following photoexcitation.}
    \label{SI_7}
\end{figure}

\begin{figure}[H]
    \centering
    \includegraphics[width=1.0\linewidth]{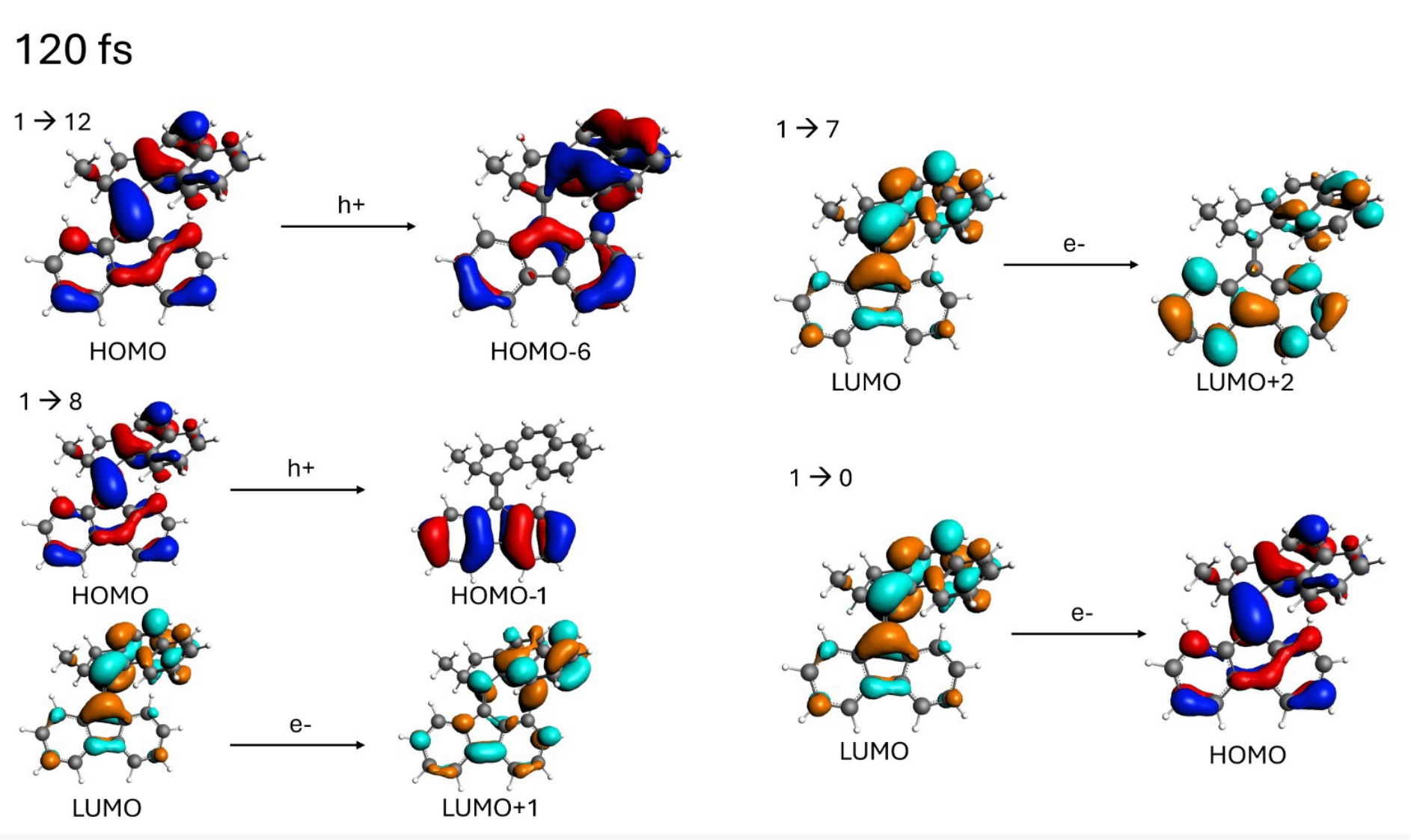}
    \caption{Isosurfaces of the frontier molecular orbitals involved in the four representative hole (h$^+$) and electron (e$^-$) transitions at 120 fs, labeled by the corresponding excitation (1 $\rightarrow$ 12, 1 $\rightarrow$ 7, 1 $\rightarrow$ 8, 1 $\rightarrow$ 0). Occupied orbitals (HOMO, HOMO-1, HOMO-6) are shown in blue and red, denoting the two opposite phases of the wavefunction; unoccupied orbitals (LUMO, LUMO+1, LUMO+2) are shown in orange and cyan, again denoting opposite phases. Arrows indicate the orbital-to-orbital character of each hole or electron transfer following photoexcitation. }
    \label{SI_8}
\end{figure}

\begin{figure}[H]
    \centering
    \includegraphics[width=1.0\linewidth]{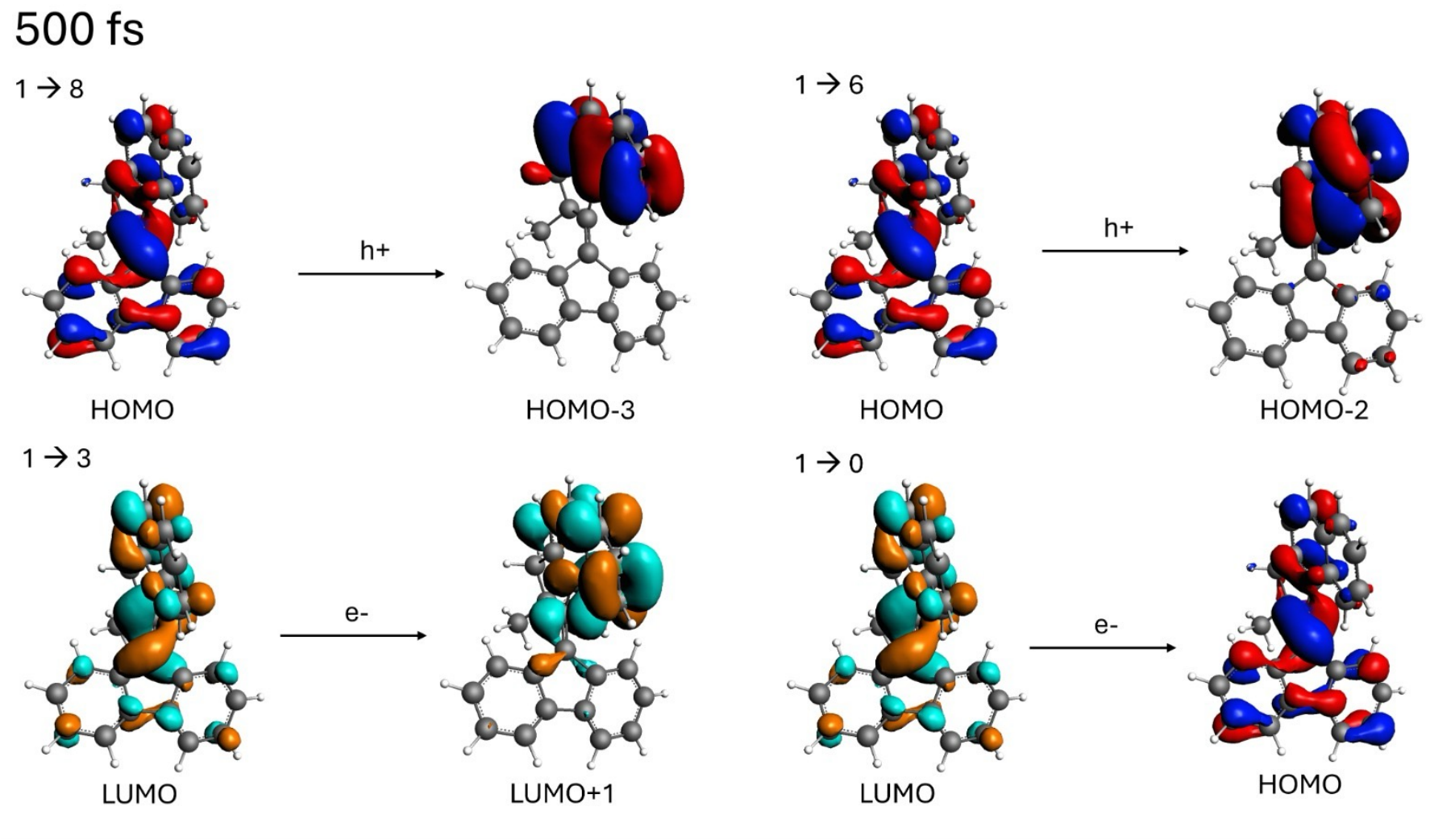}
    \caption{Isosurfaces of the frontier molecular orbitals involved in the four representative hole (h$^+$) and electron (e$^-$) transitions at 500 fs, labeled by the corresponding excitation (1 $\rightarrow$ 8, 1 $\rightarrow$ 6, 1 $\rightarrow$ 3, 1 $\rightarrow$ 0). Occupied orbitals (HOMO, HOMO-2, HOMO-3) are shown in blue and red, denoting the two opposite phases of the wavefunction; unoccupied orbitals (LUMO, LUMO+1) are shown in orange and cyan, again denoting opposite phases. Arrows indicate the orbital-to-orbital character of each hole or electron transfer following photoexcitation.}
    \label{SI_9}
\end{figure}

\begin{figure}[H]
    \centering
    \includegraphics[width=1.0\linewidth]{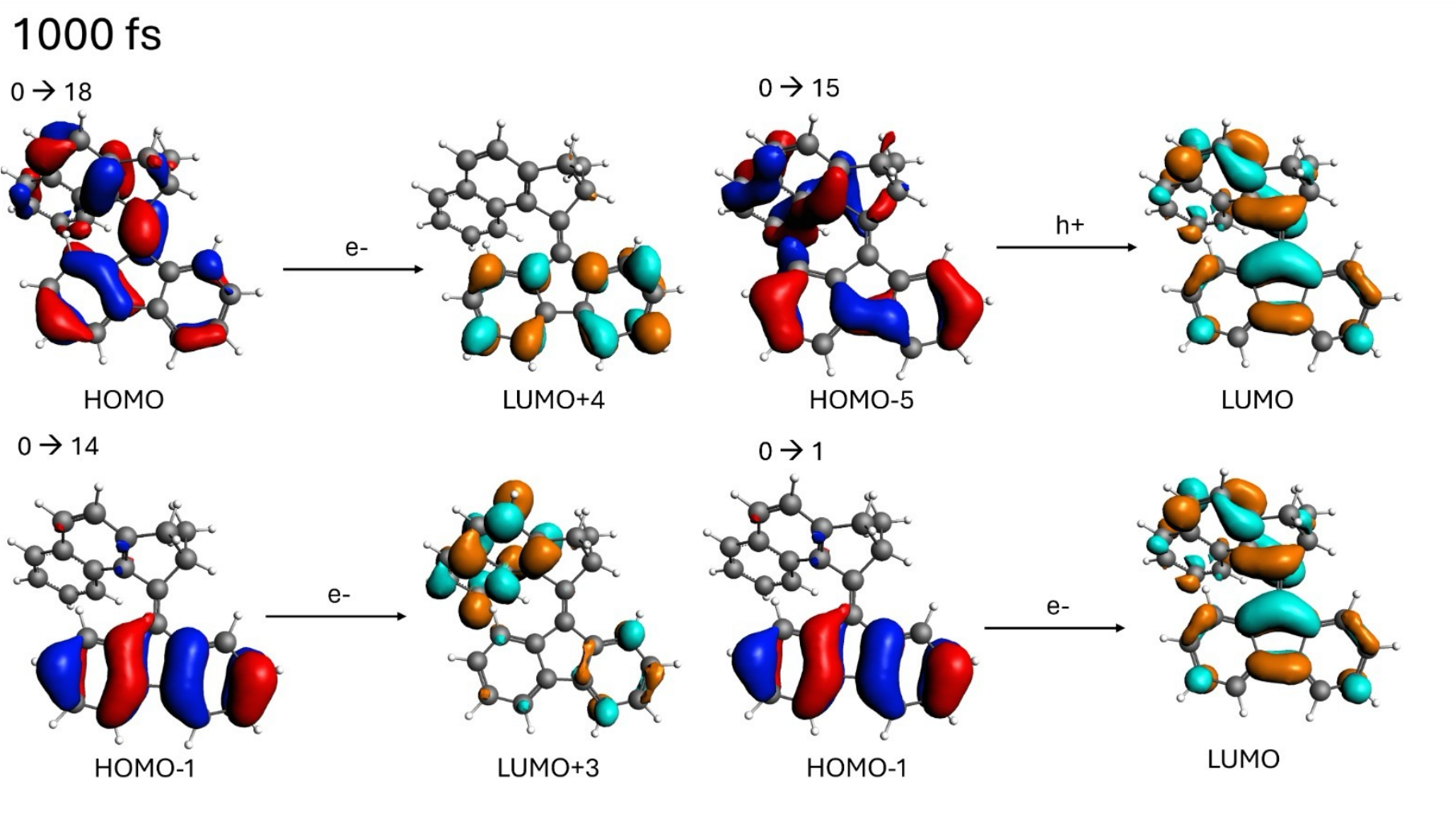}
    \caption{Isosurfaces of the frontier molecular orbitals involved in the four representative hole (h$^+$) and electron (e$^-$) transitions at 500 fs, labeled by the corresponding excitation (1 $\rightarrow$ 8, 1 $\rightarrow$ 6, 1 $\rightarrow$ 3, 1 $\rightarrow$ 0). Occupied orbitals (HOMO, HOMO-1, HOMO-5) are shown in blue and red, denoting the two opposite phases of the wavefunction; unoccupied orbitals (LUMO, LUMO+3, LUMO+4) are shown in orange and cyan, again denoting opposite phases. Arrows indicate the orbital-to-orbital character of each hole or electron transfer following photoexcitation. }
    \label{SI_10}
\end{figure}

\section{Trajectory-resolved transient absorption}

\begin{figure}[H]
    \centering
    \includegraphics[width=1.0\linewidth]{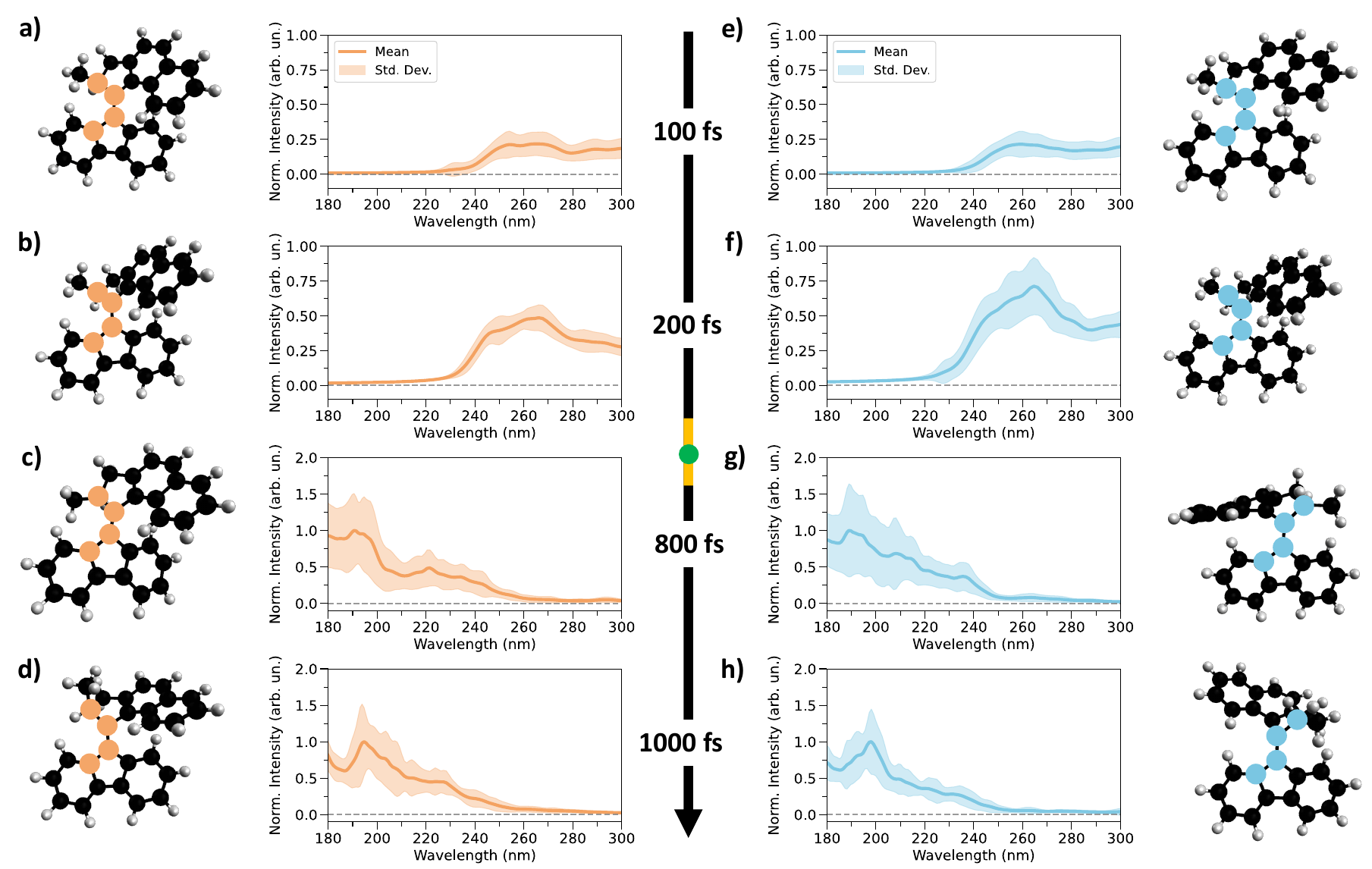}
    \caption{(a--d) Mean normalized absorption spectra (solid lines) with standard deviation (Sdt. Dev.; shaded area) computed along the trajectories associated with the P isomer (orange) at 100, 200, 800, and 1000 fs, with representative molecular structures shown on the left. (e--h) Same for the trajectories associated with the M isomer (blue), with representative structures shown on the right. Spectra are shown in the 180--300 nm range. The central timeline indicates the temporal progression of the dynamics; the green dot marks the mean S$_1$ $\rightarrow$ S$_0$ transition time and the yellow bar the corresponding standard deviation.}
    \label{SI_11}
\end{figure}

\begin{figure}[H]
    \centering
    \includegraphics[width=1.0\linewidth]{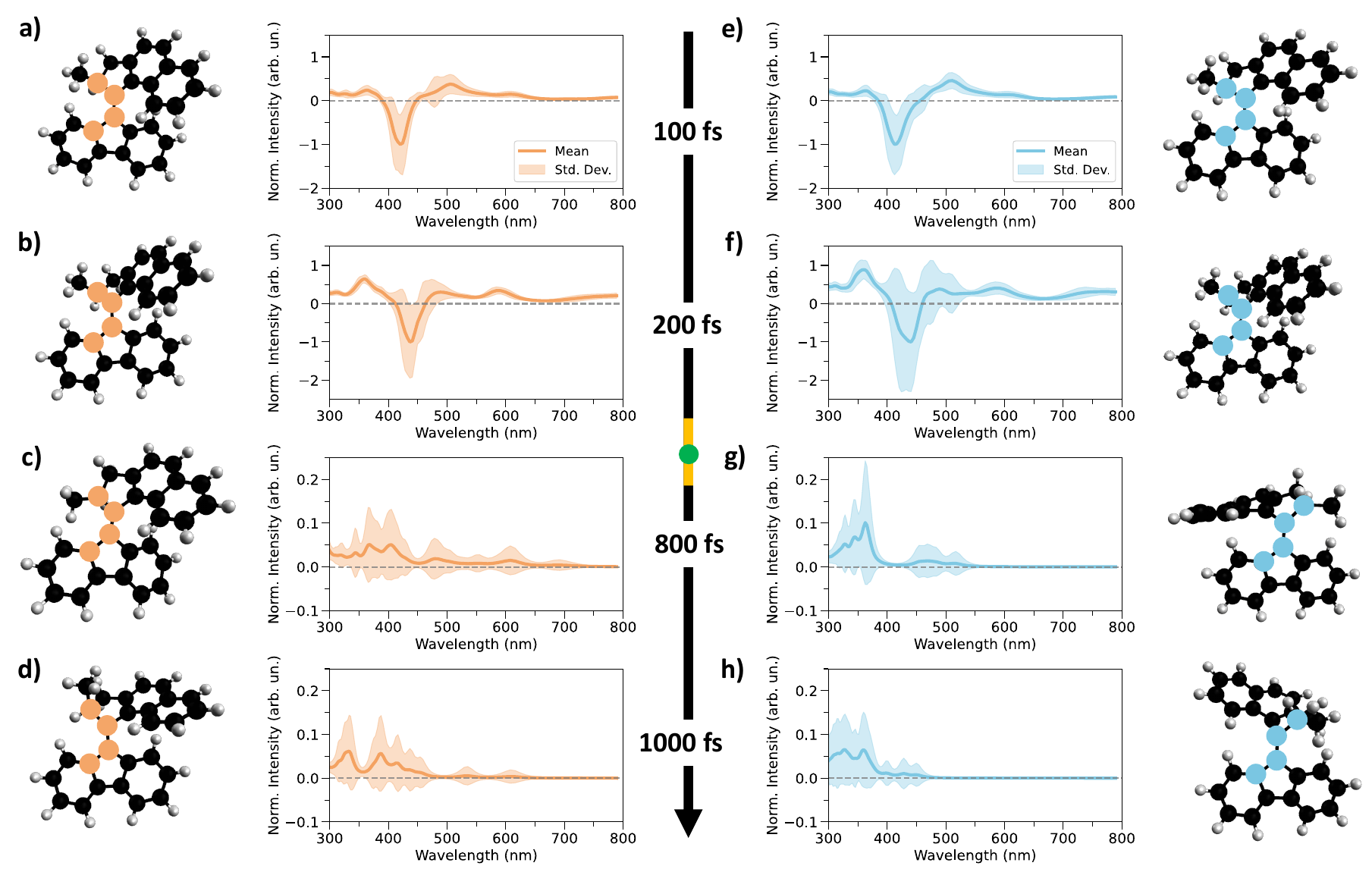}
    \caption{(a--d) Mean normalized absorption spectra (solid lines) with standard deviation (Sdt. Dev.; shaded area) computed along the trajectories associated with the P isomer (orange) at 100, 200, 800, and 1000 fs, with representative molecular structures shown on the left. (e--h) Same for the trajectories associated with the M isomer (blue), with representative structures shown on the right. Spectra are shown in the 300--800 nm range. The central timeline indicates the temporal progression of the dynamics; the green dot marks the mean S$_1$ $\rightarrow$ S$_0$ transition time and the yellow bar the corresponding standard deviation.}
    \label{SI_12}
\end{figure}

\begin{figure}[H]
    \centering
    \includegraphics[width=1.0\linewidth]{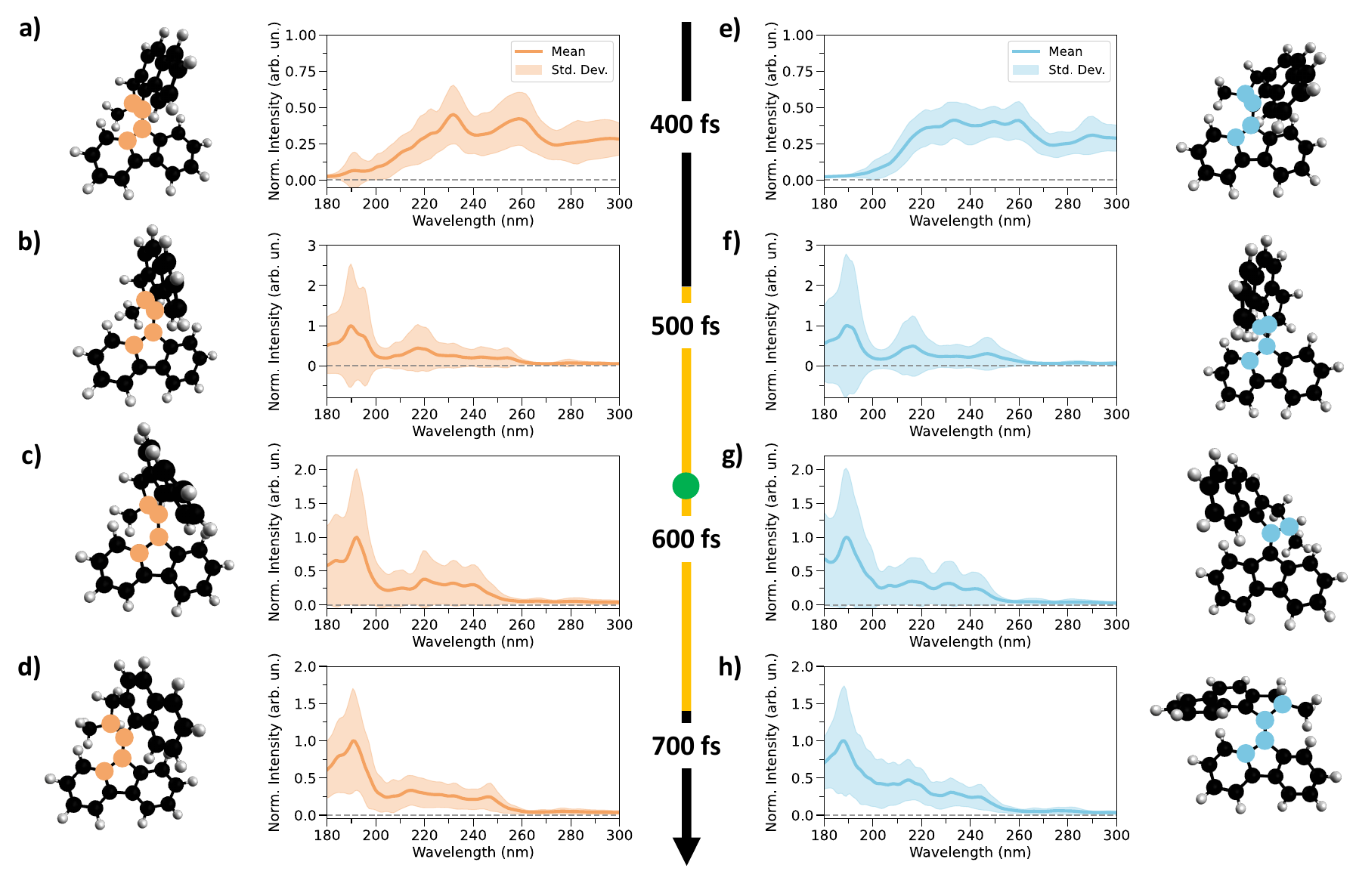}
    \caption{(a--d) Mean normalized absorption spectra (solid lines) with standard deviation (Sdt. Dev.; shaded area) computed along the trajectories associated with the P isomer (orange) at 400, 500, 600, and 700 fs, with representative molecular structures shown on the left. (e--h) Same for the trajectories associated with the M isomer (blue), with representative structures shown on the right. Spectra are shown in the 180-300 nm range. The central timeline indicates the temporal progression of the dynamics; the green dot marks the mean S$_1$ $\rightarrow$ S$_0$ transition time and the yellow bar the corresponding standard deviation.}
    \label{SI_13}
\end{figure}

\begin{figure}[H]
    \centering
    \includegraphics[width=1.0\linewidth]{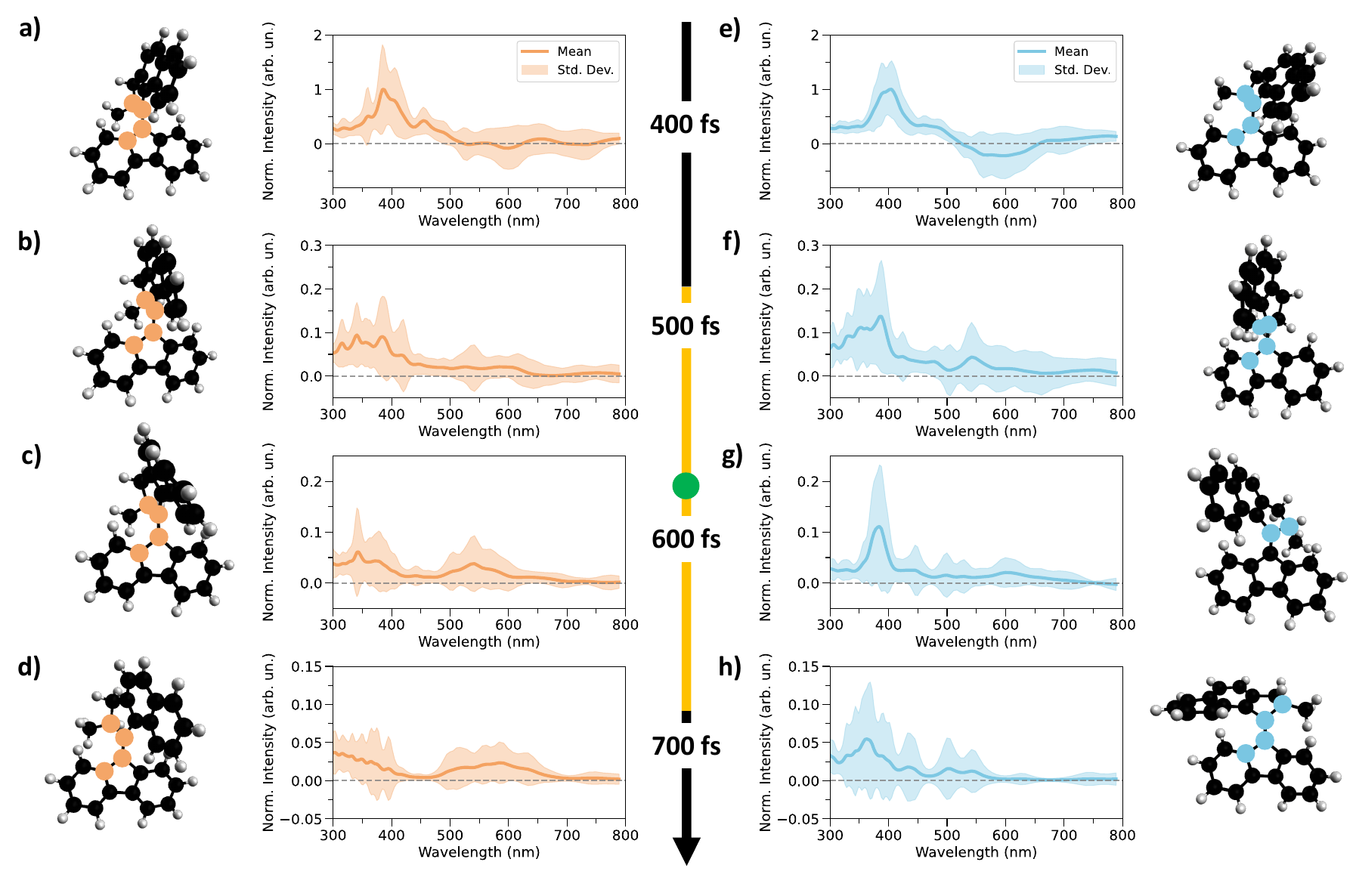}
    \caption{(a--d) Mean normalized absorption spectra (solid lines) with standard deviation (Sdt. Dev.; shaded area) computed along the trajectories associated with the P isomer (orange) at 400, 500, 600, and 700 fs, with representative molecular structures shown on the left. (e--h) Same for the trajectories associated with the M isomer (blue), with representative structures shown on the right. Spectra are shown in the 300--800 nm range. The central timeline indicates the temporal progression of the dynamics; the green dot marks the mean S$_1$ $\rightarrow$ S$_0$ transition time and the yellow bar the corresponding standard deviation.}
    \label{SI_14}
\end{figure}

As discussed in the main text, we separate the trajectory ensemble into P-bound and M-bound subsets according to the eventual photoproduct, in order to test whether the optical response carries isomer-specific information. Figures \ref{SI_11} and \ref{SI_12} show the subset-averaged absorption spectra at 100, 200, 800, and 1000 fs in the 180--300 nm and 300--800 nm ranges, respectively. Figures \ref{SI_13} and \ref{SI_14} report the same wavelength ranges across the conical intersection (CoIn) window (400--700 fs).

At early times (100 and 200 fs; panels a--e and b--f of Figures \ref{SI_11} and \ref{SI_12}), the P- and M-bound populations show nearly identical absorption profiles across both wavelength ranges, consistent with both sets of trajectories still sampling the same P-like ES region. Across the CoIn window (400--700 fs; Figures \ref{SI_13} and \ref{SI_14}), the two isomers begin to diverge geometrically along the dihedral coordinate (Figure \ref{SI_dih}), and the absorption spectra evolve substantially over this interval. Two main bands below 250 nm appear and grow in intensity (Figure \ref{SI_12}), consistent with the transfer to S$_0$, while features above 500 nm reduce to a single, weak residual positive band between 500 and 700 fs (Figure \ref{SI_14}, panels b--f, c--g, and d--h). Despite these substantial changes, the evolution proceeds in essentially the same way for both populations. At later times (800 and 1000 fs; Figure \ref{SI_11}, panels c--g and d--h), once all trajectories have decayed to the GS and the two photoisomers are maximally separated in the dihedral coordinate (Figure \ref{SI_dih}), the P and M photoproducts remain dominated by similar positive UV absorption below $\sim$250 nm, with only weak features at longer wavelengths (350--400 nm) and no response above 500 nm. The absorption profiles of the two pathways therefore overlap almost completely throughout the dynamics.

\clearpage

\clearpage

\clearpage

\section{Optical transitions in time-resolved circular dichroism}

\begin{table}[H]
\centering
\caption{Rotatory strengths computed along the nonadiabatic dynamics at 0, 120, 200, 500, and 1000 fs, within the 450–800 nm spectral window. For each selected time, we report the most relevant electronic transitions contributing to the absorption spectrum. For each transition, the average wavelength $\overline{\lambda}$ and rotatory strengths $\overline{R}$ are reported together with their standard deviations, obtained from the ensemble of trajectories. The rotatory strength is defined as $R = (\mu_x m_x + \mu_y m_y + \mu_z m_z)/3$.}
\begin{tabular}{|c|l|c|c|c|c|}
\hline
\rule{0pt}{2.8ex}\textbf{Time (fs)} & \textbf{Transition} &
\textbf{$\overline{\lambda}$ $\pm \; \sigma_\lambda$ (nm)}  & \textbf{$\overline{R} \pm \sigma_R$} \\
\hline
\multirow{4}{*}{0}
 & $|1\rangle \to |0\rangle$   & 344 $\pm$ 6&  -0.0812 $\pm$ 0.0303 \\
 & $|1\rangle \to |17\rangle$  & 516 $\pm$ 9& -0.0129 $\pm$ 0.0171 \\
 & $|1\rangle \to |14\rangle$  & 567 $\pm$ 9&  0.0463 $\pm$ 0.0396 \\
 & $|1\rangle \to |8\rangle$   & 754 $\pm$ 17&  0.0007 $\pm$ 0.0209 \\
\hline
\multirow{4}{*}{120}
 & $|1\rangle \to |0\rangle$   & 427 $\pm$ 12 &  -0.1530 $\pm$ 0.0378 \\
 & $|1\rangle \to |12\rangle$  & 501 $\pm$ 10 &  0.0300 $\pm$ 0.0529 \\
 & $|1\rangle \to |8\rangle$   & 608 $\pm$ 14 & -0.0343 $\pm$ 0.0234 \\
 & $|1\rangle \to |7\rangle$   & 768 $\pm$ 26 & -0.0113 $\pm$ 0.0080 \\
\hline
\multirow{4}{*}{200}
 & $|1\rangle \to |0\rangle$   & 445 $\pm$ 16&  -0.1703 $\pm$ 0.0307 \\
 & $|1\rangle \to |14\rangle$  & 474 $\pm$ 12&  0.0251 $\pm$ 0.0222 \\
 & $|1\rangle \to |9\rangle$   & 589 $\pm$ 14&  0.0989 $\pm$ 0.0373 \\
 & $|1\rangle \to |7\rangle$   & 744 $\pm$ 26&  0.0031 $\pm$ 0.0108 \\
\hline
\multirow{4}{*}{500}
 & $|1\rangle \to |0\rangle$   & 844 $\pm$ 285&  -0.0955 $\pm$ 0.2092 \\
 & $|1\rangle \to |8\rangle$   & 513 $\pm$ 54&  0.0148 $\pm$ 0.0207 \\
 & $|1\rangle \to |6\rangle$   & 618 $\pm$ 83&  0.0376 $\pm$ 0.0895 \\
 & $|1\rangle \to |3\rangle$   & 756 $\pm$ 125& -0.0232 $\pm$ 0.0430 \\
\hline
\multirow{5}{*}{1000}
 & $|0\rangle \to |18\rangle$  & 211 $\pm$ 5& 0.0025 $\pm$ 0.0879 \\
 & $|0\rangle \to |15\rangle$  & 218 $\pm$ 6&  0.0409 $\pm$ 0.0859 \\
 & $|0\rangle \to |14\rangle$  & 221 $\pm$ 6&  0.0205 $\pm$ 0.0739 \\
 & $|0\rangle \to |1\rangle$   & 370 $\pm$ 56&  0.0679 $\pm$ 0.1364 \\
\hline
\end{tabular}
\end{table}

\section{Trajectory-resolved time-resolved circular dichroism}

\begin{figure}[H]
    \centering
    \includegraphics[width=1.0\linewidth]{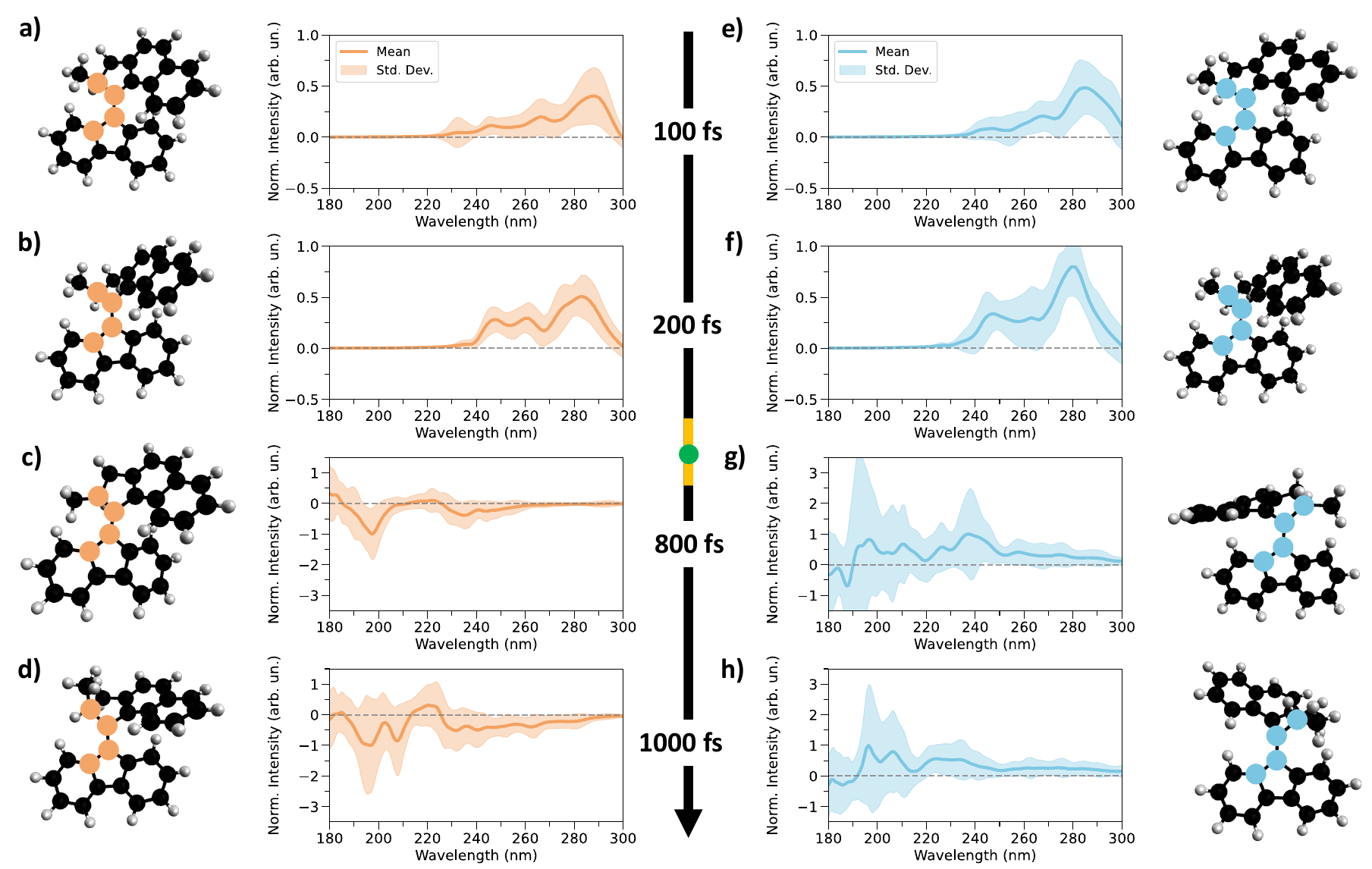}
    \caption{Trajectory-resolved time-dependent electronic circular dichroism (ECD). (a--d) Mean normalized ECD spectra (solid lines) with standard deviation (Std. Dev.; shaded area) computed along the trajectories associated with the P isomer (orange) at 100, 200, 800, and 1000 fs, with representative molecular structures shown on the left. (e--h) Same for the trajectories associated with the M isomer (blue), with representative structures shown on the right. Spectra are shown in the 180--300 nm range. The central timeline indicates the temporal progression of the dynamics; the green dot marks the mean S$_1$ $\rightarrow$ S$_0$ transition time and the yellow bar the corresponding standard deviation.}
    \label{SI_15}
\end{figure}

\begin{figure}[H]
    \centering
    \includegraphics[width=1.0\linewidth]{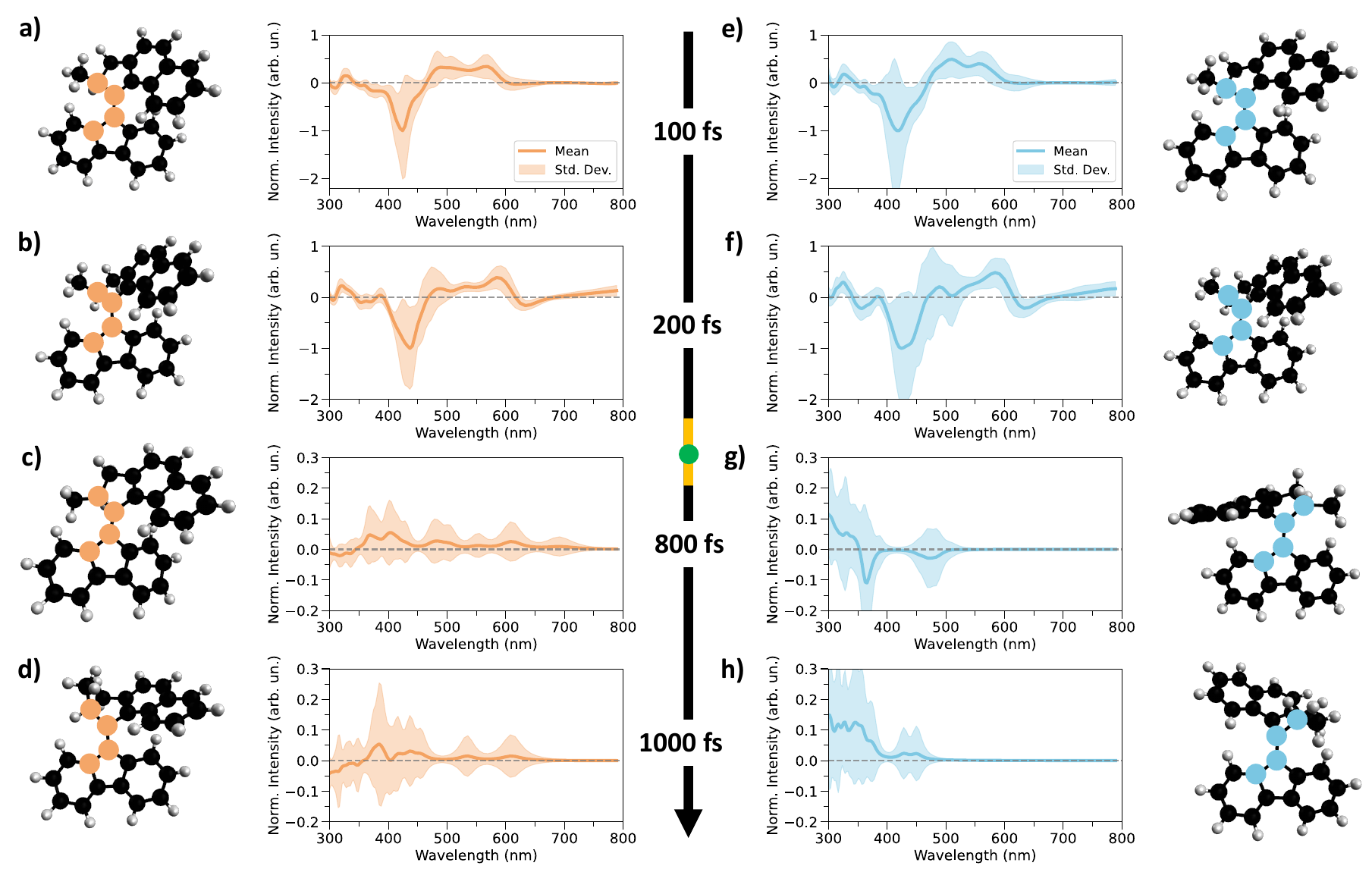}
    \caption{Trajectory-resolved time-dependent electronic circular dichroism (ECD). (a--d) Mean normalized ECD spectra (solid lines) with standard deviation (Std. Dev.; shaded area) computed along the trajectories associated with the P isomer (orange) at 100, 200, 800, and 1000 fs, with representative molecular structures shown on the left. (e--h) Same for the trajectories associated with the M isomer (blue), with representative structures shown on the right. Spectra are shown in the 300--800 nm range. The central timeline indicates the temporal progression of the dynamics; the green dot marks the mean S$_1$ $\rightarrow$ S$_0$ transition time and the yellow bar the corresponding standard deviation.}
    \label{SI_16}
\end{figure}

\begin{figure}[H]
    \centering
    \includegraphics[width=1.0\linewidth]{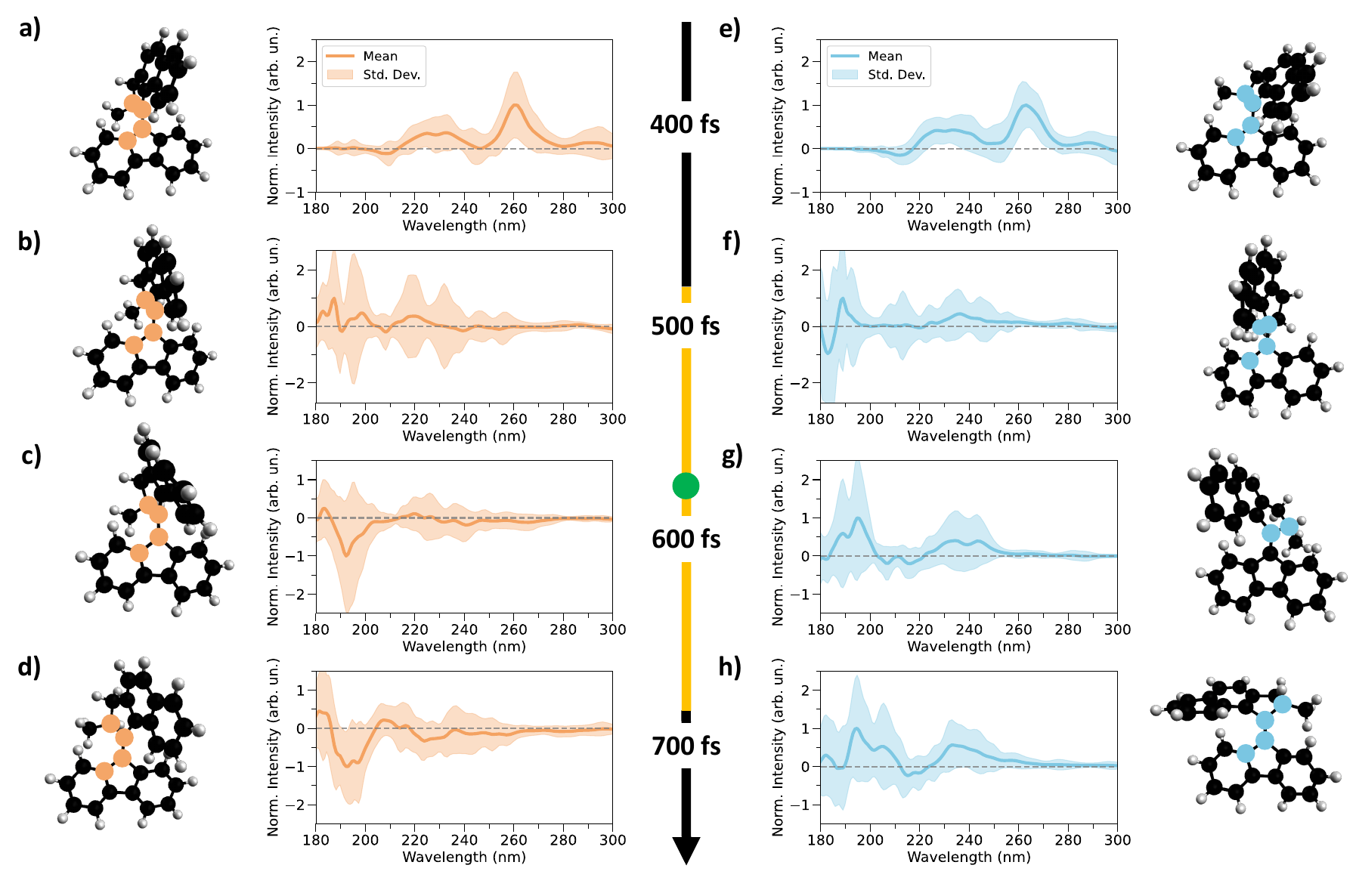}
    \caption{Trajectory-resolved time-dependent electronic circular dichroism (ECD). (a--d) Mean normalized ECD spectra (solid lines) with standard deviation (Std. Dev.; shaded area) computed along the trajectories associated with the P isomer (orange) at 400, 500, 600, and 700 fs, with representative molecular structures shown on the left. (e--h) Same for the trajectories associated with the M isomer (blue), with representative structures shown on the right. Spectra are shown in the 180--300 nm range. The central timeline indicates the temporal progression of the dynamics; the green dot marks the mean S$_1$ $\rightarrow$ S$_0$ transition time and the yellow bar the corresponding standard deviation.}
    \label{SI_17}
\end{figure}

This section reports the subset-averaged TRCD spectra in full detail, complementing the analysis in the main text. Figures \ref{SI_15} and \ref{SI_16} report the trajectory-resolved ECD spectra at 100, 200, 800, and 1000 fs in the 180--300 nm and 300--800 nm regions, respectively, providing a comparison between the early ES response and the late GS photoproduct signatures. The intermediate CoIn window is shown separately in Figure \ref{SI_17}, where the 180--300 nm spectral evolution is followed from 400 to 700 fs, complementing the corresponding 300--800 nm analysis reported in the main text in Figure 4.

At early times, the two subsets show nearly identical features across both wavelength ranges (Figures \ref{SI_15} and \ref{SI_16}), consistent with the trajectories still sampling the same P-like ES region. After passage through the CoIn, distinct chiroptical signatures emerge for the two photoproduct pathways. While the 300--800 nm response is discussed in the main text, the same differentiation is also visible in the UV region, particularly at 600 and 700 fs (Figure \ref{SI_17}). The P-bound trajectories show a strong negative feature around 190 nm and a weaker one centered around 230 nm, while the M-bound trajectories develop two positive bands around $\sim$195 nm and 240 nm. These signatures are consistent with the emergence of the characteristic ground-state ECD features of the two isomers\cite{vicario2005controlling}, which become fully established at later delays (800 and 1000 fs; Figure \ref{SI_15}, panels c--g and d--h) once the ensemble has decayed to the GS. Residual discrepancies with the experimental ground-state ECD remain, reflecting the vibrationally hot character of the photoproducts at these delays. As discussed in the main text, the P-bound spectra retain a weak positive feature between 500 and 700 nm at 800 fs (Figure \ref{SI_16}, panels c--g), whereas the M-bound response is quenched, providing the chiral discrimination between the two photoproducts.

\clearpage

\bibliography{bib}

\end{document}